%% file: main.tex
\RequirePackage{fix-cm}

\documentclass[smallextended]{svjour3}
\smartqed

\usepackage[T1]{fontenc}
\usepackage{cite}
\usepackage{algorithmic}
\usepackage{graphicx}
\usepackage{textcomp}
\usepackage{booktabs}
\usepackage{footmisc}
\usepackage{longtable}

\usepackage{microtype}
\usepackage{siunitx}
\usepackage{minibox,soul}
\usepackage{xspace}
\usepackage[inline]{enumitem}
\usepackage[xspace]{ellipsis} 
\usepackage[margin= 1.0in]{geometry}
\usepackage[section]{placeins}

\usepackage{rotating}
\usepackage{longtable}
\usepackage{multirow}
\usepackage{soul}
\usepackage{hyperref}
\usepackage{framed}
\usepackage{lscape}
\usepackage{amsmath,scalerel}

\usepackage{pifont}
\usepackage{selinput}
\usepackage{rotating}
\usepackage{capt-of}
\usepackage{afterpage}
\usepackage{pdflscape}
\usepackage{inputenc}
\usepackage{amssymb}

\usepackage[table,xcdraw]{xcolor}
\usepackage{colortbl}

\usepackage{tabularx}

\definecolor{Mygray}{rgb}{0.90,0.90,0.90}

\UseRawInputEncoding

\begin{document}

\title{Incubation and Beyond: A Comparative Analysis of ASF Projects Sustainability Impacts on Software Quality}

\author{Adam Alami \and Steffan Klockmann \and Lasse Rehder S\o rensen \and Ra{\'u}l Pardo \and Johan Lin\r{a}ker}

\institute{A. Alami \at
              University of Southern Denmark\\
              M{\ae}rsk Mc-Kinney M{\o}ller Institute \\
              Alsion 2  \\
              S\o nderborg, Denmark  \\
              \email{adal@mmmi.sdu.dk}
              \and
           S. Klockmann\at
               Aalborg University\\
               Aalborg, Denmark\\
               \email{klockmann.steffan@gmail.com}
                \and
           L. Rehder S\o rensen\at
               Aalborg University\\
               Aalborg, Denmark\\
               \email{rehderfris@gmail.com}
                \and
           R. Pardo\at
              IT University of Copenhagen\\
              Denmark\\
              \email{raup@itu.dk}
              \and
           J. Lin\r{a}ker\at
              RISE Research Institutes of Sweden \\
              Sweden \\
              \email{johan.linaker@ri.se}
}

\date{Received: date / Accepted: date}

\maketitle

\begin{abstract}

\textbf{Context:} Free and Open Source Software (FOSS) communities' sustainability, meaning to remain operational without signs of weakening or interruptions to its development, is fundamental for the resilience and continuity of society's digital infrastructure. Many of our digital services and products either leverage or entirely rely on FOSS somewhere in their software stack. However, FOSS sustainability is a multifaceted concept, and the impact of its decline on community products is less known.

\textbf{Objective:} In this study, we sought to understand how the different aspects of FOSS sustainability impact software quality from a life-cycle perspective. Specifically, we aim to investigate whether and how support and incubation of FOSS projects or bypassing incubation correlate with software quality outcomes.

\textbf{Method:} We selected 342 FOSS projects from the Apache Software Foundation that have either graduated, retired, or bypassed their incubator program. We used 16 sustainability metrics to examine their impact on eight software quality metrics (defect density, code coverage, medium risk complexity, very high risk complexity, number of large files, very large function size count, code duplication percentage, and most complex function LOC). We analyzed the decline in the sustainability metrics and their relationship with our selected software quality metrics using Bayesian data analysis, which incorporates probability distributions to represent the regression coefficients and intercepts.

\textbf{Results:} We found that our selected sustainability metrics exhibit distinct relationships with software quality across different project trajectories. Graduated projects demonstrated the strongest sustainability-software quality (SWQ) relationship, both during and post-incubation. In contrast, retired projects exhibited weaker relationships, despite receiving similar governance support. Bypassed projects, while not outperforming graduated ones, showed comparable sustainability-SWQ relationships.

\textbf{Conclusion:} While structured incubation strengthens sustainability and SWQ in graduated projects, retired projects struggle to maintain strong sustainability-SWQ relationships, indicating that additional factors internal and specific to projects influence sustainability. This effect was evident among bypassed projects; their self-reliant sustainability practices yielded stronger sustainability-software quality (SWQ) compared to the retired ones.


\end{abstract}

\input{introduction.tex}
\input{related.tex}

\input{sustainability}
\input{swq}
\input{hypotheses}
\input{methods.tex}
\input{findings.tex}
\input{discussion.tex}
\input{validity}

\input{conclusion.tex}

\section{Statements and Declarations}

\paragraph{Data Availability Statements.}\label{sec:replication}

\noindent We made this study's data and other artifacts available \href{https://figshare.com/s/64a074506dca255fa733}{here.}\footnote{\url{https://figshare.com/s/64a074506dca255fa733}}

\paragraph{Conflict of Interest.} The authors declared that they have no conflict of interest.

\paragraph{Ethical approval}
Not applicable.

\paragraph{Informed consent}
Not applicable.

\paragraph{Author Contributions}
Adam Alami was the lead author, performed a majority of the writing, and participated actively in the scoping, design, implementation, analysis, writing, and reviewing of the paper.
Steffan Klockmann and Lasse Rehder participated actively in the scoping and design discussions of the study, and as well as data collection.
Raul Pardo performed the majority of the data analysis, generated plots and figures, developed and assumed the ownership of the scripts development and tools integration, and participated actively in the scoping, design, implementation, analysis, writing, and reviewing of the paper. 
Johan Lin\r{a}ker participated actively in the scoping, design, implementation, analysis, and reviewing of the paper.

\input{acks.tex}

\bibliographystyle{spmpsci}
\bibliography{references}

\end{document}

%% file: introduction.tex
\section{Introduction}\label{sec:introduction}

Free and Open Source Software (FOSS) has become integral to the continuity of today's digital infrastructure \cite{crowston2008free}. Its adoption in many digital services and applications solidifies it as a core public asset. However, FOSS's sustainability remains a concern \cite{fang2009understanding}. Recent vulnerabilities in widely adopted FOSS packages and the disruptions caused, e.g., OpenSSL's ``Heartbleed'' (2014), Equifax Breach (2017), NPM packages ``event-stream'' (2018), and ``colors'' (2022), accentuate the need for FOSS communities to remain sustainable \cite{stuanciulescu2022code}. This includes their ability to draw in resources to maintain and innovate their products \cite{chengalur2010sustainability}.

FOSS ``sustainability'' and community ``health'' are used interchangeably in the literature \cite{linaaker2022characterize}, but they mean the same concept. Chengalur-Smith et al. define sustainability as the ability of the community to maintain its activities and continue the development of its products over time \cite{chengalur2010sustainability}. Similarly, Lin\r{a}ker et al. define community health as its resilience to remain operational without signs of weakening or interruptions to its development \cite{linaaker2022characterize}.

The sustainability of FOSS is essential for not only enabling broader access to technology and innovation but also facilitating access to freely available software to various demographics irrespective of their economic and geographic conditions \cite{alami2024free}. Nonetheless, community sustainability impact on product's properties, such as quality and security, is less understood \cite{alami2024free,stuanciulescu2022code}. Work on the topic, which examined the relationship between sustainability and software quality (SWQ), emphasized one sustainability indicator at a time. For example, Foucault et al. investigated the relationship between developer turnover and modules' quality \cite{foucault2015impact}, Wang et al. investigated the effort distributions of elite developers on productivity and quality \cite{wang2020unveiling}, and Ghapanchi investigated the relationship between sustainability and the community capacity to fix bugs, add new features, and manage releases \cite{ghapanchi2015predicting}. Stuanciulescu et al. found that graduating from the Apache Software Foundation incubator (ASFI) is reliant on code quality, among other factors \cite{stuanciulescu2022code}. However, this study assumed that the decision to graduate or retired is a proxy for sustainability, which is not always the case. The decision is a board decision process subject to criteria not always sustainability-related and influenced by the performance of competing projects. In sum, these studies offer either a fragmented understanding of the relationship between SWQ and sustainability or use alternate benchmarks for sustainability not inline with indicators reported in the literature \cite{linaaker2022characterize}.

To overcome these limitations, in our previous work \cite{alami2024free}, we used 16 sustainability indicators as advised in Lin{\aa}ker et al.'s framework for assessing FOSS sustainability \cite{linaaker2022characterize}. Although our findings show no evidence of a consistent positive or negative impact of our selected sustainability metrics on software quality \cite{alami2024free}, we treated all ASFI projects equally, regardless of their developmental trajectory, e.g., retired, graduated, during incubation, or post-incubation. This approach may have obscured potential and meaningful variations in the sustainability-SWQ relationship across different trajectories. For instance, we still lack clarity on whether projects that graduated perform differently compared to those that bypassed incubation or those that retired. 

Projects following ASF incubation benefit from structured governance and mentorship, whereas bypassed projects rely on self-imposed discipline. Do these differing sustainability conditions lead to consistent software quality outcomes? Likewise, a retired project may not suffer from poor initial quality but could experience degradation over time due to weak sustainability practices. Treating all projects uniformly obscures how different sustainability conditions shape software quality across these trajectories. This categorization---\textbf{Retired}, \textbf{Graduated during incubation}, \textbf{Graduated post-incubation}, and \textbf{Bypassed incubation}---may allow us to identify patterns that were previously indiscernible, potentially refining our understanding of the impact of community sustainability on FOSS software quality. 

Unlike previous studies that treated ASF projects uniformly, our study introduces a project trajectory-based categorization to reveal sustainability-quality patterns that have remained unknown. To this end, we ask:

\medskip 

\noindent \textbf{\emph{RQ:}} \emph{How does the sustainability of ASF communities affect their software quality across various developmental trajectories---Retired, Graduated during incubation, Graduated post-incubation, and Bypassed incubation?}

\medskip

The ASF ecosystem offers a unique opportunity to examine the relationship between FOSS sustainability and software quality through the lens of project trajectories. The ecosystem has projects at various levels of development and sustainability statuses. For example, projects like ``Kafka'' \cite{Kafka} and ``Spark'' \cite{Spark} successfully graduated from incubation and became widely adopted, benefiting from strong governance structures and sustained community engagement. In contrast, projects like ``Wave'' \cite{Wave} were retired despite initial interest, highlighting potential sustainability challenges. Meanwhile, projects such as ``Pulsar'' \cite{Pulsar} and ``Iceberg'' \cite{Iceberg} bypassed incubation, demonstrating that alternative pathways can lead to successful FOSS projects, possibly with different sustainability strategies. Examining how these distinct trajectories and their sustainability patterns influence software quality allows us to uncover whether certain sustainability indicators are more critical for supporting SWQ and whether ASF’s incubation process provides measurable benefits in sustaining projects, and if any cascading impact on software quality.

We selected projects from the ASF foundation including both those who followed and bypassed the incubation process. Our data covered projects' activities in the period from March 2009 to April 2023. We mined project data from inception (first activity we found in repertories) to the last activity, at the time we downloaded the data. Our dataset includes 342 projects after applying our selection criteria. We mined the data using Git and Perceval, see section \ref{sec:methods} for further details. For data analysis, we used Bayesian statistics.

We found that project trajectories influence the sustainability-SWQ relationship. Graduated projects show stronger code quality, both during and post-incubation. The strength of this relationship becomes stronger post-incubation. Retired projects exhibited the weakest sustainability-SWQ relationship among the trajectories. We also found that this relationship is stronger in graduated projects compared to bypassed projects.

We contribute by providing a novel empirical investigation into the role of project trajectories in moderating the relationship between FOSS sustainability and software quality. Unlike prior work that treated sustainability as a uniform concept and static across projects and their development, our study differentiates between graduated, post-graduation, bypassed, and retired projects, revealing distinct sustainability-quality patterns. Specifically, we contribute:

\begin{itemize}

    \item \textbf{A trajectory-based framework for analyzing the sustainability-quality relationship.} We introduce a categorization of FOSS projects based on their developmental trajectory—graduated, post-graduation, bypassed, and retired—to systematically assess how sustainability influences software quality across different governance and development pathways. While this categorization may not generalize to all FOSS projects, our findings demonstrate that FOSS sustainability is not a uniform concept. Instead, it should be analyzed through both developmental and temporal lenses, to recognize its dynamics across time and trajectories.

    \item \textbf{Empirical evidence of sustainability's impact on software quality across trajectories.}  Our Bayesian analysis shows that graduated projects benefit the most from sustainability practices, exhibiting stronger effects on software quality both during and after incubation. In contrast, retired projects demonstrate a weaker relationship between sustainability and software quality, despite receiving similar mentoring and governance support from ASF. While this suggests that their sustainability indicators are inherently weaker, it also indicates that ASF governance alone is insufficient to sustain projects.
    
\end{itemize}

We commence this paper with a summary of earlier work on the impact of FOSS sustainability on software quality in Sect. \ref{sec:related}. Then, in Sect. \ref{sec:sustainability}, we explain how we measure FOSS sustainability and software quality in Sect. \ref{sec:swq}. In Sect. \ref{sec:hypotheses}, we present our hypotheses. In Sect. \ref{sec:methods}, we describe our research methods, and we share our findings in Sect. \ref{sec:findings}. We discuss the implications of our work in Sect. \ref{sec:discussion}. We highlight threats to validity in Sect. \ref{sec:validity}. Finally, Sect. \ref{sec:conclusion} brings us to conclusion.

%% file: related.tex
\section{Related Work}\label{sec:related}

Although few studies addressed the sustainability-SWQ relationship \cite{stuanciulescu2022code,foucault2015impact,wang2020unveiling}, they remain inconclusive. This outcome is partly due to the divergence in approaches taken by different studies, and the relationship is simply not linear as assumed and cannot be evaluated by linear methods alone \cite{alami2024free}. For example, while St{\u{a}}nciulescu, et al. used retirement and graduation as a surrogate indicator for sustainability \cite{stuanciulescu2022code}, Foucault et al. emphasized developer turnover as a single sustainability metric \cite{foucault2015impact}. In addition, while Foucault et al. found negative impacts of turnover and defect density \cite{foucault2015impact}, Alami et al. report no statistical evidence of positive or negative impacts \cite{alami2024free}.

Foucault et al. opted to investigate a single sustainability indicator, i.e., developer turnover, impact on bug density, and fixes per module of five open source projects \cite{foucault2015impact}. They found that while external turnover (leaving the community) has a negative influence on defect density and fixes, internal turnover (moving to another module within the community) has no effect \cite{foucault2015impact}. Wang et al. used a different sustainability indicator, i.e., effort distributions of elite or core developers, to examine its impact on quality \cite{wang2020unveiling}. They found that the number of new defects discovered monthly in each project is inversely related to the effort invested by elite developers in organizational and support tasks \cite{wang2020unveiling}. This finding was echoed in previous studies. For example, Mockus et al. reported that the top 15 contributors (out of 388 overall) contributed more than 83\% of change requests and 66\% of the resolution of issue reports in AFS communities \cite{mockus2002two}.

St{\u{a}}nciulescu, et al. investigated the relationship between project graduation and code, process quality of ASFI projects \cite{stuanciulescu2022code}. They found that retired projects had somewhat greater cyclomatic complexity; however, their retirement is not linked with defects, complexity, or technical debt. Other quality metrics such as file size, function size, and function complexity seem to be the most unfavorable to projects sustainability \cite{stuanciulescu2022code}. In our previous work \cite{alami2024free}, we pursued similar objective; however, our conceptualization of sustainability was different and our hypotheses were aligned to investigate the direct relationship between FOSS sustainability and software quality. To this end, we used 16 sustainability metrics reported in Lin{\aa}ker et al.'s framework for assessing the sustainability of FOSS  \cite{linaaker2022characterize} and analyzed how they impact software quality, using eight software quality metrics (defect density, code coverage, medium risk complexity, very high risk complexity, number of large files, very large function size count, code duplication percentage, and most complex function LOC). We found no evidence of a consistent positive or negative impact of our selected sustainability metrics on software quality metrics \cite{alami2024free}.

While some studies posited that when FOSS sustainability declines, some quality indicators may deteriorate \cite{foucault2015impact,wang2020unveiling}, St{\u{a}}nciulescu, et al. and Alami et al. found no link with some code quality characteristics and project retirement and no statistical evidence of positive or negative impacts, respectively \cite{stuanciulescu2022code,alami2024free}. Yet, an important factor remains unexamined: the developmental trajectory of FOSS projects. A targeted analysis into the various trajectories---graduates, bypasses incubation, or retires---may introduce key distinctions into the sustainability-SWQ relationship. For instance, projects that successfully graduate may demonstrate a stronger sustainability-SWQ relationship during and post incubation period due to higher organizational maturity and resource attraction, e.g., more contributors. Bypassed projects could also show unique sustainability dynamics and subsequent impact due to their alternative pathway. Lastly, retired projects may show challenges tied to specific sustainability indicators compared to graduated and bypassed projects.

%% file: sustainability.tex
\section{Measuring FOSS sustainability}\label{sec:sustainability}

To assess the sustainability of FOSS communities in our sample, we adopted the framework developed by Lin{\aa}ker et al. \cite{linaaker2022characterize}. The framework provides an overview of 107 sustainability indicators organized into 15 themes (see Tbl.~\ref{tbl:originalframework}). The framework is based on the synthesis of 146 related publications on FOSS sustainability. It covers themes spanning communication, popularity, stability, and technical activity. This coverage allows for a holistic assessment of FOSS sustainability and its complex and multi-layered nature. The choice is a departure from the use of singular measures in previous studies, e.g., \cite{foucault2015impact,wang2020unveiling}. This methodological choice also allowed us to cover multiple indicators in our analysis, thereby contributing to a nuanced analysis of the sustainability-SWQ relationship.

\begin{table*}[th!]
    
    \footnotesize
    \vspace{2mm}
    
    \renewcommand\arraystretch{1.0}
    \caption{High level overview of the 15 themes with the corresponding number of identified sustainability metrics across the Actor, Software, and Orchestrational perspectives, from Lin{\aa}ker et al.' framework \cite{linaaker2022characterize}. Our adoption is presented in table~\ref{tbl:framework}.}
    
    \label{tbl:originalframework}
  
    \begin{tabular}{p{3cm}p{10.3cm}c}
    
    \toprule
      \textbf{Theme} & \textbf{Description} & \textbf{\# Metrics}\\ \hline

        \multicolumn{2}{l}{\textbf{Actors-perspective}} & \\ \midrule
        Communication  & 
        This theme captures interactions within and outside the FOSS community~\cite{linaaker2022characterize} & 4\\ 
      
        Culture & 
        This theme defines the community inclusiveness and openness towards potential actors~\cite{linaaker2022characterize}.& 6 \\ 
      
        Finance & 
        This theme captures the financial support and stability received by the actors maintaining and contributing to the community~\cite{linaaker2022characterize}. & 2 \\ 
      
        Diversity & 
        ``Describes the OSS project's or its overarching ecosystem's ability to be receptive to diversity and self-renew itself''~\cite{linaaker2022characterize}. & 5 \\ 
      
        Popularity & 
        This theme describes the general external interest in the community and its ecosystem~\cite{linaaker2022characterize}. & 7 \\ 
      
        Stability & 
        This theme captures the resilience and robustness of the community in terms of population, i.e., attracting and sustaining a steady flow of contributors~\cite{linaaker2022characterize}. & 12 \\ 
      
        Technical activity & 
        This tehme describes the overall technical activity within the community~\cite{linaaker2022characterize} 5 \\ \midrule

        \multicolumn{2}{l}{\textbf{Software-perspective}} & \\ \midrule
      
        Documentation & 
        \textit{``Describes the quality of general and technical documentation.''}~\cite{linaaker2022characterize} & 6 \\ 
      
        Development process & 
        This theme describes the quality and maturity of the development processes and practices~\cite{linaaker2022characterize}. & 7 \\ 
      
        License & 
        \textit{``Highlights how license choices and related practices may affect the popularity and attractiveness of an OSS project, both for commercial actors and individuals.''}~\cite{linaaker2022characterize} & 4 \\ 
      
        General factors & 
        This theme describes the attractiveness of the community based on its general technical features~\cite{linaaker2022characterize}. & 7 \\ 
      
        Scaffolding & 
        This theme defines the accessibility and robustness of the development and the communication infrastructure used by the community to facilitates a highly collaborative environment~\cite{linaaker2022characterize}. & 5 \\ 
      
        Security & 
        This theme captures the community resilience and ability to address current and future vulnerabilities in its code base~\cite{linaaker2022characterize}. & 6 \\ 
      
        Technical quality & 
        This theme defines community quality practices~\cite{linaaker2022characterize}. & 10 \\ \midrule

        \multicolumn{2}{l}{\textbf{Orchestration-perspective}} & \\ \midrule
      
        Orchestration & 
        This theme captures governance structures and the quality of the leadership in the community~\cite{linaaker2022characterize}. & 9 \\
    
    \bottomrule
      
  \end{tabular}

\end{table*}

Prior to data mining and analysis, we reduced the number of themes and corresponding metrics to those more frequently cited in the literature and most accessible for mining in an MSR study. This rationalization exercise helped us to emphasize the most relevant and practical metrics. For example, within the communication theme, ``response time'' is referenced more than ``visibility'' (the community's social media presence); whereas the latter is cited four times, the former is reported ten times \cite{linaaker2022characterize}. Furthermore, some themes and their associated parameters cannot be sourced using our selected methods (see Sect. \ref{sec:methods}). For example, the theme ``culture'' and its corresponding parameters, such as ``conflicts,'' can only be sourced through qualitative methods. Nevertheless, our reduced instance of the framework (Tbl. \ref{tbl:framework}) remains extensive, with four themes and 16 parameters.

\begin{table*}[th!]
    
    \footnotesize
    \vspace{2mm}
    
    \renewcommand\arraystretch{1.0}
    \caption{FOSS sustainability parameters adopted from Lin{\aa}ker et al. \cite{linaaker2022characterize}. ``Dev'' refers to development. Detailed computation methods for each metric are available in Sect. \ref{sec:methods}.}
    \label{tbl:framework}
  
    \begin{tabular}{p{2cm}p{9mm}p{1.8cm}p{9.5cm}}
    
    \toprule
      \textbf{Theme} & \textbf{ID} & \parbox{2cm}{\textbf{Sustainability}\\\textbf{parameter}\vspace{1mm}} & \textbf{Definition}\\ \hline
      
        \multirow{7}{2.5cm}{Communication} & \multirow{2}{1cm}{COM-1} & \multirow{2}{3.1cm}{Response time} & The time elapsed between when a community member posted a comment/question and when he or she got a reply/response \cite{linaaker2022characterize}. \textit{Computation:} We computed the response time by calculating the average time taken for the first comment to appear on an issue across all projects. \\ \cline{2-4}

        & \multirow{2}{1cm}{COM-2} & \multirow{2}{3.1cm}{Frequency of communication} & Frequency in FOSS project communication, such as the number of problems opened or comments posted in a certain time frame \cite{linaaker2022characterize}. \textit{Computation:} The total number of comments and issues posted across the projects within a specific timeframe, quantifying the communication frequency. \\ \hline
        
        \multirow{3}{2.5cm}{Popularity} & POP-1 & \parbox{2cm}{\vspace{1mm}Project\\popularity} & The projects' overall popularity, as reflected in forks, stars, and watchers \cite{linaaker2022characterize}. textit{Computation: Project popularity was computed by summing the number of forks, stars, and watchers for each project, reflecting external interest and engagement.} \\ \hline

        \multirow{30}{2.5cm}{Stability} & \multirow{1}{1cm}{STA-1} & \multirow{1}{3.1cm}{Age} & The age of the community is the duration in years from inception to current date \cite{linaaker2022characterize}. \textit{Computation:} The number of years from the project's inception to the current date, representing the age of the project.\\ \cline{2-4}
        
        & \multirow{2}{1cm}{STA-2} & \multirow{2}{3.1cm}{Attrition} & Attrition is the gradual reduction in contributors. This means that contributors are leaving faster than they are newcomers to the community \cite{linaaker2022characterize}. \textit{Computation:} Attrition was measured by the cumulative decrease in the number of commits over specified twelve-week periods to capture the reduction in active contributions.\\ \cline{2-4}

        & STA-3 & Forks & The number of project's forks. \cite{linaaker2022characterize}. \textit{Computation:} The number of forks represents how many times the project has been forked by other users\\ \cline{2-4}

        & \multirow{1}{1cm}{STA-4} & \multirow{1}{3.1cm}{Growth} & FOSS project growth and progress, as well as overall technological activity. \cite{linaaker2022characterize}. \textit{Computation:} Growth was quantified by the cumulative increase in the number of PRs submitted over twelve-week intervals, indicating active development and enhancement.\\ \cline{2-4}

        & \multirow{2}{1cm}{STA-5} & \multirow{2}{3.1cm}{Knowledge concentration} & Distribution of contributions and expertise to specific persons or groups within an FOSS project, usually measured and explained using a community's bus- or truck factor \cite{linaaker2022characterize}. \textit{Computation:} We used the truck factor method \cite{avelino2016novel} to compute this metric, this metric assesses how concentrated the project knowledge is among a few contributors.\\ \cline{2-4}

        & \multirow{1}{1cm}{STA-6} & \multirow{1}{3.1cm}{Life-cycle stage}  & The stage in the life cycle a FOSS project currently resides in. For example, growth, or dormancy \cite{linaaker2022characterize}. \textit{Computation:} We adopted Valiev et al.'s definition of dormancy, ``having very little or no development activity after some time'' \cite{valiev2018ecosystem}, to compute this metric (see Sect. \ref{sec:methods} for further details).\\ \cline{2-4}

        & \multirow{2}{1cm}{STA-7} & \multirow{2}{3.1cm}{Retention} & The capacity of a FOSS community to keep its contributors active in the community for an extended length of time \cite{linaaker2022characterize}. \textit{Computation:} Retention was measured by the annual increase in the number of active contributors.\\ \cline{2-4}

        & \multirow{1}{1cm}{STA-8} & \multirow{1}{3.1cm}{Size} & The size of the FOSS project in terms of users and developers in any specific point in time \cite{linaaker2022characterize}. \textit{Computation:} We computed the total the number of unique contributors who have engaged in at least one commit, PR, issue, or comment, indicating the community size.\\ \cline{2-4}

        & \multirow{2}{1cm}{STA-9} & \multirow{2}{3.1cm}{Turnover} & The number of contributors and maintainers who depart a FOSS community in a given time frame \cite{linaaker2022characterize}. \textit{Computation:} Turnover is computed by counting contributors who authored commits but have been inactive for the past six months, showing the rate of contributor departure.\\ \hline

        \multirow{20}{2.5cm}{\parbox{2cm}{Technical\\activity}} & \multirow{2}{1cm}{TEC-1} & \multirow{2}{3.1cm}{\parbox{2cm}{\vspace{1mm}Contributors'\\dev.activity}} & This value relates to the degree of non-maintainer contributions, such as development and technical writing. Specifically, it measures the total technical output excluding activities performed by the maintainers \cite{linaaker2022characterize}. \textit{Computation:} We computed this metric by counting all commits from contributors who are not maintainers to capture involvement of general community members in technical activities.\\ \cline{2-4}

        & \multirow{2}{1cm}{TEC-2} & \multirow{2}{3.1cm}{Efficiency} & The effectiveness and convenience with which a FOSS project manages and advances development, such as accepting and evaluating issues and pull-requests \cite{linaaker2022characterize}. \textit{Computation:} We measured the average time from PR creation to its closure to reflect the project's operational effectiveness.\\ \cline{2-4}

        & \multirow{2}{1cm}{TEC-3} & \multirow{2}{3.1cm}{\parbox{2cm}{\vspace{1mm}Non-code\\contributions}} & Technical activity of specifically non-code-contributions, i.e. documentation, community management, answering questions etc. \cite{linaaker2022characterize}. \textit{Computation:} It is the counts of commits that affect non-code files like documentation to capture contributions beyond code.\\ \cline{2-4}

        & \multirow{2}{1cm}{TEC-4} & \multirow{2}{3.1cm}{\parbox{2cm}{\vspace{1mm}Overall dev.\\activity}} & Technical activity overall by the community, for example, development and technical writing. I.e. including both the maintainers' and contributors' work \cite{linaaker2022characterize}. \textit{Computation:} This metric is the count of commits of coding files (other than ``txt'' and ``md'').\\ 
    
    \bottomrule
      
  \end{tabular}

\end{table*}

%% file: swq.tex
\section{Measuring software quality}\label{sec:swq}

\noindent The ISO/IEC 25010 framework defines software quality as ``the degree to which the system satisfies the stated and implied needs of its various stakeholders and thus provides value'' \cite{iso2011}. The ISO/IEC 25010 model also proposes additional non-functional characteristics to include in the definition of software quality. These include ``functional suitability'', ``performance efficiency'', ``compatibility'', ``usability'', ``reliability'', ``security'', ``maintainability'', and ``portability'' \cite{iso2011}. 

Software quality is a complex concept that is challenging to define and measure \cite{kitchenham1996software,alami2022scrum}. For example, while ISO/IEC 25010 provides a comprehensive framework to define SWQ \cite{iso2011}, not all the proposed attributes are relevant to all circumstances and projects. For example, the model attributes like ``usability'' and ``portability'' may be less relevant for a backend system designed purely for internal use within a single environment.

In this study, constrained by the inherent limitations of MSR techniques, we confined the definition of SWQ to defect density and code quality (see Tbl. \ref{tbl:swquality}), in line with our previous work \cite{alami2024free} and similar studies, i.e., Stuanciulescu et al. \cite{stuanciulescu2022code}. This alignment allowed us to compare our findings with previous work (see Sect. \ref{sec:discussion}). In addition, our SWQ metrics exceed the scope of similar studies, e.g., \cite{khomh2012faster,ray2014large,wang2020unveiling}. For example, Wang et al. used a precise but narrow definition, i.e., \textit{``the number of bugs found during a project-month''} \cite{wang2020unveiling}.

Table \ref{tbl:swquality} documents our SWQ metrics, which incorporate a broader array of quality attributes. We include defect density and also several code quality metrics, such as code coverage and various complexity metrics (e.g., McCabe index). Our metrics cover both internal attributes, influenced by developers, and external attributes, reflecting end users expectations of quality \cite{alami2024free}. Our approach, to some extent, aligns with the ISO/IEC 25010 definition (\cite{iso2011}). For example, our data include defects reported in issue trackers covering a broad array of defect types, including functionality, usability, and performance issues.

\begin{table*}[th!]
    
    \footnotesize
    \vspace{2mm}
    
    \renewcommand\arraystretch{1.0}
    \caption{Software Quality Metrics Adopted in this Study}
    \label{tbl:swquality}
  
    \begin{tabular}{p{1.5cm}p{2cm}p{11cm}}
    
    \toprule
      \textbf{ID} & \textbf{Metric} & \textbf{Definition}\\ \hline
      
        \multirow{2}{1.5cm}{\textbf{SWQ-1}} & \multirow{2}{2cm}{Defect density} & The defect density is defined as the number of confirmed bugs in a program or module during a given time period of operation or development divided by the total size of the software or module (KLOC, size of the release, or per development day) \cite{shah2013software}.\\ \hline
        
        \multirow{4}{1.5cm}{\textbf{SWQ-2}} & \multirow{4}{2cm}{Code quality} & Code quality attributes ``not visible to the end-users'', i.e., ``internal'' and ``structural'' qualities \cite{alami2022scrum}. This metric encompasses code coverage ((Number of lines or statements tested / Total number of lines or statements) x 100\%) (\textbf{SWQ-2.1}), medium risk complexity (McCabe index between 11-25) (\textbf{SWQ-2.2}), very high risk complexity (McCabe index $>$50) (\textbf{SWQ-2.3}), number of very large files ($>$1000 SLOC) (\textbf{SWQ-2.4}), Very Large Function Size Count (\textbf{SWQ-2.5}), Code Duplication Percentage (\textbf{SWQ-2.6}), and Most Complex Function LOC (\textbf{SWQ-2.7}) \cite{stuanciulescu2022code}.\\ 
    
    \bottomrule
      
  \end{tabular}

\end{table*}

%% file: hypotheses.tex
\section{Hypotheses Development}\label{sec:hypotheses}

Based on the Lin{\aa}ker et al. \cite{linaaker2022characterize} framework, we propose that FOSS sustainability can be measured using four themes and their associated indicators: \textbf{communication}, \textbf{popularity}, \textbf{stability}, and \textbf{technical activity} \cite{linaaker2022characterize}. 

\textbf{Communication.} Lin{\aa}ker et al.'s themes \cite{linaaker2022characterize} are well anchored in the literature. For example, Wang et al. suggest that the efficiency of communication to contributors is critical to community survival \cite{wang2012survival}. Communication quality is defined as the combination of quality (degree of information and accuracy) \cite{linaaker2022characterize}, response time or responsiveness \cite{jiang2019metrics}, and frequency of communicating \cite{shaikh2019selecting}.

Studies reported that a decline in communication quality and efficiency lead to dissatisfied contributors \cite{steinmacher2019overcoming,wang2012survival,guizani2021long} and, consequently a decline in community sustainability \cite{linaaker2022characterize}. Quality and efficient communication ensures that contributors feel appreciated, engaged, and encouraged to continue contributing \cite{van2017health}. When communication deteriorates, delays increases, which may impact the software development efficiency and quality. Under the communication theme, we adopted two sustainability parameters from Lin{\aa}ker et al.'s work \cite{linaaker2022characterize}: \emph{response time} (\textbf{COM-1}), and \emph{frequency of communication} (\textbf{COM-2}). 

\textbf{Popularity.} This theme captures the overall external interest in the FOSS project \cite{linaaker2022characterize}. Zhou and Mockus argue that a positive community outlook, reflected in project's popularity serves to diminish the likelihood of leaving the community \cite{zhou2012make}. In similar vein, Osman et al. argue that sustainable projects are more popular, and are likely to be actively developed and maintained \cite{osman2021health}. Borges et al. also report that three out of four end-users consider the project's star rating, although they add that such metrics should be used cautiously \cite{borges2016understanding}. For example, a fast increase in the number of stars could be the result of promotion on social sites and may have little to do with an active and sustainable development. Acknowledging this limitation, including it as a sustainability metric is an opportunity to test its reliability as a sustainability indicator \cite{alami2024free}. 

\textbf{Stability.} The stability of the FOSS community is defined by its ability to attract and maintain a consistent population of contributors over an extended period of time \cite{linaaker2022characterize}. The literature reported growth, retention, attrition, turnover of contributors \cite{foucault2015impact}, the concentration of knowledge (i.e., ``truck factor'') \cite{wang2020unveiling}, and the community size as sustainability indicators \cite{linaaker2022characterize}. 

The potential reduction in contributors to support and maintain the community software, may lead to a decline in the efficiency of the development process. For example, a shortage in contributors can result in reduced testing, and a decreased frequency of updates and bug fixes. As suggested by Lin{\aa}ker et al. \cite{linaaker2022characterize}, under this theme, we selected nine sustainability parameters: \emph{age} (\textbf{STA-1}), \emph{attrition} (\textbf{STA-2}), \emph{forks} (\textbf{STA-3}), \emph{growth} (\textbf{STA-4}), \emph{knowledge concentration} (\textbf{STA-5}), \emph{life-cycle stage} (\textbf{STA-6}), \emph{retention} (\textbf{STA-7}), \emph{size} (\textbf{STA-8}), and \emph{turnover} (\textbf{SAT9}).

\textbf{Technical activity.} Lin{\aa}ker and colleagues define this theme as the technical production to advance and enhance the community software products \cite{linaaker2022characterize}. Maintaining a high level of technical activities can be reflected in the continuous addition of features and the resolution of bugs \cite{linaaker2022characterize}. Midha and Palvia argue that community ``technical success'' relies on the level of developer activity \cite{midha2012factors}. They reported that a higher number of contributors is positively correlated with higher technical activities \cite{midha2012factors}.  Mockus et al. also add that technical production is necessary beyond the core team \cite{mockus2002two}.

It is argued that when technical activity is reduced, the project may experience stagnation and decreased in novel functionality and bugs resolution \cite{alami2024free}. Under this theme, we adopted four sustainability indicators: \emph{contributors' development activity} (\textbf{TEC-1}), \emph{efficiency} (\textbf{TEC-2}), \emph{non-code contributions} (\textbf{TEC-3}), and \emph{overall development activity} (\textbf{TEC-4}).

Drawing on these conceptual bases, we develop hypotheses to test how these themes' sustainability indicators impact SWQ. In brief, we hypothesize that FOSS sustainability’s influence on software quality is contingent on project trajectory: graduated projects benefit more from sustainability practices than retired projects, while bypassed projects may demonstrate equal or stronger relationships than formally incubated ones. Subsequently, SWQ may vary in its responsiveness to sustainability indicators depending on a project's trajectory, being more sensitive to sustainability variations in retired projects compared to bypassed and graduated projects. We hypothesize:

\medskip

\noindent\textit{\textbf{H1:} Graduated projects exhibit a stronger positive relationship between sustainability and software quality metrics compared to retired projects during the incubation period.}

\medskip

\noindent\textit{\textbf{H2:} Post incubation, graduated projects exhibit a stronger positive relationship between sustainability and software quality metrics compared to their time during incubation.}

\medskip

\noindent\textit{\textbf{H3:} Bypassed projects exhibit an equal or stronger positive relationship between sustainability and software quality metrics compared to graduated projects post incubation.}

\medskip

\noindent\textit{\textbf{H4:} The relationship between sustainability metrics and software quality is moderated by project trajectory, with bypassed and graduated projects showing a stronger relationship than retired projects.}

\medskip

The evaluation these hypotheses will allow us to understand how these sustainability themes (communication, popularity, stability, and technical activity), and their associated parameters, impact SWQ. These insights will help uncover whether certain sustainability dimensions play a more dominant role in shaping SWQ depending on a project's developmental trajectory. Specifically, we aim to determine whether sustainability factors that drive software quality in well-supported and structured projects (e.g., graduated ones) differ from those in self-sustaining (bypassed) or declining (retired) projects. By distinguishing between trajectories, we can assess whether sustainability influences different software quality attributes (e.g., defect density) to varying degrees. Ultimately, this investigation contributes to a more nuanced understanding of how sustainability mechanisms interact with project lifecycles to shape SWQ outcomes.

%% file: methods.tex
\section{Methods}\label{sec:methods}

\subsection{Data collection \& computation}

Continuing from our earlier work~\cite{alami2024free}, we focus our investigation of FOSS projects within the confines of the ASF. Newly adopted projects typically join ASFs Incubator program, while some who exhibit sustainability and maturity from the eyes of the ASF, may bypass the incubator program and directly join the general confines of foundation. The projects, either incubated or bypassed, go through a development trajectory, aiming to promote maturity and resilience. Although not all projects follow a similar trajectory, the sample is a unique opportunity to examine the sustainability-SWQ relationship across various phases of community development. In addition, ASF projects offer a diverse range of projects spanning various domains, from infrastructure and data processing to web development and artificial intelligence \cite{alami2024free}. This diversity allowed us to examine the sustainability-SWQ relationship in various software development domains.

We used ASF's \href{https://projects.apache.org/projects.html}{project list}\footnote{\url{https://projects.apache.org/projects.html}} to select projects in our sample. To delineate projects that bypassed incubation from those who did, we used ASF's \href{https://incubator.apache.org/projects/#current}{Podlings list}\footnote{\url{https://incubator.apache.org/projects/\#current}}. Both lists are reliable sources for project's statuses. 
Our sample contains 398 projects. However, some projects did not meet our selection criteria. We excluded projects without PRs in GitHub or Jira, empty issue trackers, non-GitHub/Jira issue hosting, and unavailable defect labels (essential for \textbf{SWQ-1} (\emph{defect density}) computation). Our final sample contains 342 projects. We mined and computed our data from scratch; we did not re-use previous work dataset~\cite{alami2024free}.

To mine projects' data, we used the git tool to clone all projects and \href{https://github.com/chaoss/grimoirelab-perceval}{Perceval}\footnote{\url{https://github.com/chaoss/grimoirelab-perceval}} to aggregate data components and attributes such as commits, issues, and PRs, stars, watchers, forks, and repository size in KB. We also developed additional Python scripts to clone repositories from GitHub; extract commits from the cloned repositories; download issues, PRs, repository information from GitHub; and Jira issues. We used \href{https://www.sokrates.dev/}{Sokrates}\footnote{\url{https://www.sokrates.dev/}} to extract \textbf{SWQ-2} (\emph{code quality}) metrics (see Tbl. \ref{tbl:swquality}). Sokrates is a static code analysis, which computes a variety of code quality metrics, such as cyclomatic complexity, maintainability index, and code duplication. The size of the collected dataset is 255 GB uncompressed and 71 GB in a compressed tarball.

We compute metrics from a ``start date'' to an ``end date'', which depend on the projects' trajectory under evaluation. There are four cases depending on the trajectory of projects:
\begin{enumerate*}[label=\roman*)]
\item for bypassed projects, we compute metrics from the inception date of the repository to the most recent date of project activity;
\item for retired projects, from the inception date of the repository to the date of retirement;
\item for graduated projects, from the inception date of the repository to the date of graduation;
\item finally, for post-incubation graduated projects, we compute metrics from graduation date to the most recent date of project activity.
\end{enumerate*}
In what follows, we describe how each metric was calculated. Below we use ``start date'' and ``end date'' to refer to the aforementioned dates; which map to the corresponding values depending on the trajectory of projects for which they are computed.

For \textbf{COM-1} (\emph{response time}), we calculated the average time it takes for the first comment to be posted in an issue. For \textbf{COM-2} (\emph{frequency of communication}), we calculated the number of comments in all issues plus the total number of issues. We opted to use issue instead of PR comments because we premise that they capture better community engagement \cite{alami2024free}. PR comments are specific to development activities; on the other hand, issues report software errors, enhancements, and new features, making them better representations of community engagement with external and internal stakeholders \cite{alami2024free}. For \textbf{POP-1} (\emph{project popularity}), we counted the total number of forks, stars, and watchers combined.

To accurately capture project behavior during incubation, we measured project's age, \textbf{STA-1} (\emph{age}), of graduated during incubation and retired projects from their start date until graduation or retirement, respectively, rather than using their current age. This ensures that our analysis reflects their actual age at the time the behavior occurred, rather than including years outside the incubation period that could introduce noise. For graduated post-incubation projects, we instead calculated age from their inception date, capturing their entire lifespan beyond incubation. Similarly, for bypassed projects, age was measured from inception to the present. This distinction allows us to analyze incubation-phase behavior separately from long-term project evolution.

For \textbf{STA-2} (\emph{attrition}), we computed the cumulative decrease in the number of commits, in periods of twelve weeks, from ``start date'' to ``end date''. \textbf{STA-3} (\emph{forks}) is the total number of a project's forks. We used the cumulative increase in the number of PRs in periods of twelve weeks to compute \textbf{STA-4} (\emph{growth}).

\textbf{STA-5} (\emph{knowledge concentration}), we used Avelino et al.'s definition of the ``truck factor'' \cite{avelino2016novel}. They define it as the minimum number of contributors who need to depart a project before it becomes compromised \cite{avelino2016novel}. To compute this metric, we use a tool\footnote{\url{https://github.com/aserg-ufmg/Truck-Factor}} developed and made available by the same researchers \cite{avelino2016novel}. To compute \textbf{STA-6} (\emph{life-cycle stage}), we adopted Valiev et al.'s definition of dormancy, ``having very little or no development activity after some time'' \cite{valiev2018ecosystem}. We computed the average number of commits per month in the last twelve months from ``end date'' of each project. If the number is below zero, then the project is considered dormant.

For retention, we used Wang et al.'s method to determine ``active'' contributors. When project's contributors perform a task within a 90-day period that requires write permission (i.e., code changes), their active status is renewed for another 3 months, starting from the month when the task was performed \cite{wang2020unveiling}. We computed \textbf{STA-7} (\emph{retention}) by calculating the cumulative total of annual increases in the number of active contributors. For \textbf{STA-8} (\emph{size}), we counted the number of contributors who have engaged in at least one commit, PR, issue, or issue comment. \textbf{STA-9} (\emph{turnover}) is the total number of contributors who authored commits and have been inactive in the last six-month from our designated ``end date''.

To compute \textbf{TEC-1} (\emph{contributors' development activity}), we counted the total number of commits made by non-maintainers, contributors who have not merged any PR yet. \textbf{TEC-2} (\emph{efficiency}) is the average time elapsed from PR creation until merger or closure. For \textbf{TEC-3} (\emph{non-code contributions}), we counted the total number of commits of files that are not related to programming code, i.e., ``txt'' and ``md'' files. \textbf{TEC-4} (\emph{overall development activity}) captures coding contributions, the count of commits of coding files other than ``txt'' and ``md.''

The \textbf{SWQ-1} (\emph{defect density}) is the ratio of defects, labeled as ``defects'' in the issues list, to the size of the project measured in kilobytes. To compute this metric, we manually inspected the issues for each project and extracted their defect labels. We found these labels in the issue tracker of the projects: ``kind:bug,'' ``bug,'' ``kind/bug,'' ``type:bug,'' ``type/bug,'' ``Bug,'' ``issue: bug,'' and ``type: bug.'' We used theses labels to identify defects. For \textbf{SWQ-2} (\emph{code quality}) metrics (see Tbl. \ref{tbl:swquality}), we used Sokrates.


\newcommand{\Issues}{\ensuremath{I}}
\newcommand{\Comments}{\ensuremath{C}}
\newcommand{\Forks}{\ensuremath{F}}
\newcommand{\Stars}{\ensuremath{S}}
\newcommand{\Watchers}{\ensuremath{W}}
\renewcommand{\time}{\ensuremath{\mathit{time}}} 
\renewcommand{\year}{\ensuremath{\mathit{year}}} 
\newcommand{\repo}{\ensuremath{\mathit{r}}}
\newcommand{\Commits}{\ensuremath{CM}}
\newcommand{\PullRequests}{\ensuremath{\mathit{PR}}}
\newcommand{\period}{\ensuremath{\mathit{period}}}
\newcommand{\periodinit}{\ensuremath{\mathit{init}}}
\newcommand{\periodend}{\ensuremath{\mathit{end}}}
\newcommand{\periodfreq}{\ensuremath{\mathit{freq}}}
\newcommand{\contributors}{\ensuremath{\mathit{contributors}}}
\newcommand{\activecontributors}{\ensuremath{\contributors_{\mathit{act}}}}
\newcommand{\contributor}{\ensuremath{\mathit{contr}}}
\newcommand{\commit}{\ensuremath{\mathit{cmt}}}
\newcommand{\mergers}{\ensuremath{\mathit{mergers}}}
\newcommand{\mergedclosed}{\ensuremath{\time_{\mathit{merged/closed}}}}
\newcommand{\pullrequest}{\ensuremath{\mathit{pr}}}
\newcommand{\files}{\ensuremath{\mathit{files}}}
\newcommand{\file}{\ensuremath{\mathit{f}}}
\newcommand{\Docs}{\ensuremath{\mathit{Docs}}}

\newcommand{\Decrements}{\ensuremath{\mathit{Decs}}}
\newcommand{\Increments}{\ensuremath{\mathit{Incs}}}
\newcommand{\NumberCommitsPeriod}{\ensuremath{\mathit{NC}}}

Tbl. \ref{tbl:smetrics} outlines the formulas we use to compute our selected sustainability parameters. In what follows, we describe the notation employed in the table. We use $\Issues$ to represent the set of issues within a project. 
$\Comments_i$ denotes the set of comments in an issue $i \in \Issues$, with $c^j_i \in \Comments_i$ specifying a concrete comment under that issue; $C \triangleq \cup_{i \in I} C_i$ encompasses all comments across project issues. 
The superscript $j \in \mathbb{N}$ in a comment $c^j_i$ indicates the chronological order among the comments in an issue, for example, $c^1_i$ is the first comment on issue $i$ and $c^3_i$ for the third comment on the same issue.
The $\time$ function retrieves creation time for issues and comments. The size of a set $S$ is denoted as $|S|$. The quantities $|\Forks|$, $|\Stars|$, and $|\Watchers|$ indicate the number of forks, stars, and watchers, respectively. 
$\Commits^t$ denotes a set of commits within a time period $t$; for example, $\Commits^{2023\text{-}05}$ denotes commits in May 2023, while $\Commits$ spans all project commits. 
The function $\period(\periodinit, \periodend, \periodfreq)$ yields a set of time periods within the start and end dates and a frequency (in weeks); for example, $\period(\text{2020-01-01},\text{2021-01-01},4)$ gives all 4-week periods from 2020-01-01 to 2021-01-01. Given a time period $t$, we use $t_\periodinit$ and $t_\periodend$ to refer to the beginning and end dates of the time period. $\PullRequests^t$ refers to the set of PRs in time period $t$. 
We use \mergedclosed\ to retrieve the time when a PR was merged or closed. 
Given the start/end date of a project, the function $\year$ returns the year of the date, e.g., $\year(\text{2020-01-01}) = 2020$. 
The function $\contributors$ takes as input a set of either commits, pull requests, issues, or comments and returns the set of contributors who created the commits, pull requests, issues, or comments, respectively. Similarly, \activecontributors\ returns the set of active contributors in a given a set of commits. Furthermore, given a contributor $\contributor$ and a time period $t$, we use $\Commits^t_\contributor$ to denote the commits of \contributor\ during time period $t$. The function $\mergers$ takes as input a set of pull requests and returns the set of contributors who merged them. Given a commit $\commit$, the function $\files(\commit)$ returns all the files modified in $\commit$. Finally, $\Docs$ denotes the set of documentation files in a project.

\begin{table*}[t]
    
    \footnotesize
    \vspace{2mm}
    
    \renewcommand\arraystretch{1.0}
    \setlength\extrarowheight{3mm}
    \caption{FOSS sustainability parameters computation methods}
    \label{tbl:smetrics}
  
    \begin{tabular}{m{1.5cm}m{4.0cm}m{9.5cm}@{}}
    
      \toprule
      \textbf{ID} 
      & \textbf{Sustainability parameter} 
      & \textbf{Computation method}
      \\ \midrule
      
      \textbf{COM-1} 
      & Response time 
      & $\sum_{i \in I} (\time(c^1_i) - \time(i)) / |I| $ 
      \\ 

      \textbf{COM-2} 
      & Frequency of communication 
      & $|\Comments| + |\Issues|$
      \\ 
        
      \textbf{POP-1}
      & Project popularity
      & $|\Forks| + |\Stars| + |\Watchers|$
      \\ 

      \textbf{STA-1}
      & Age
      & $\year(\periodend) - \year(\periodinit)$
      \\ 
        
      \textbf{STA-2}
      & Attrition
      & $ \sum_{d \in D} d \text{ with } D = \{ |\Commits^t| - |\Commits^{t+1}| \mid \Commits^j \in \Commits, t + 1, t \in \period(\periodinit,\periodend,\periodfreq) \}$
      \\ 

      \textbf{STA-3}
      & Forks
      & $|\Forks|$
      \\ 

      \textbf{STA-4}
      & Growth
      & $ \sum_{i \in \Increments} i \text{ with } \Increments \triangleq \{ |\PullRequests^{t+1}| - |\PullRequests^t|  \mid  \PullRequests^j \in \PullRequests. t + 1, t \in \period(\periodinit,\periodend,\periodfreq) \}$
      \\ 

      \textbf{STA-5}
      & Knowledge concentration
      & Algorithm 1 in \cite{avelino2016novel}
      \\ 

      \textbf{STA-6}
      & Life-cycle stage
      & $   \sum_{n \in \NumberCommitsPeriod} n / |\NumberCommitsPeriod| \text{ with } \NumberCommitsPeriod \triangleq \{ |\mathit{CM}^t \in \mathit{CM}|  \mid  t \in \period(\periodinit - \textit{1 year}, \periodend, 4) \} < 1$
      \\ 

      \textbf{STA-7}
      & Retention
      & $   \sum_{i \in \Increments} i \text{ with } \Increments \triangleq \{ |\activecontributors(\Commits^{t+1})| - |\activecontributors(\Commits^t)|  \mid  \Commits^j \in \Commits, t + 1, t \in \period(\periodinit,\periodend,\periodfreq)  \}$
      \\ 

      \textbf{STA-8}
      & Size
      & $|\contributors(\Commits) \cup \contributors(\PullRequests) \cup \contributors(I) \cup \contributors(C)|$
      \\ 

      \textbf{STA-9}
      & Turnover
      & $|\{ \contributor \in \contributors(\Commits^t) \mid \exists \commit \in \Commits^t_{\contributor} \cdot t_\periodend - \time(\commit) < \textit{6 months} \}|$
      \\ 

      \textbf{TEC-1}
      & Contributors' dev. activity
      & $|\{\contributor \in \contributors(\Commits) \mid \contributor \not \in \mergers(\PullRequests) \}|$
      \\ 

      \textbf{TEC-2}
      & Efficiency
      & $\sum_{t \in T} t/|\PullRequests| \text{ with } T = \{\mergedclosed(\pullrequest) - \time(\pullrequest) \mid \pullrequest \in \PullRequests\}$
      \\ 

      \textbf{TEC-3}
      & Non-code contributions
      & $|\{ \commit \in \Commits \mid \exists \file \in \files(\commit) \cdot \file \in \Docs \}|$
      \\ 

      \textbf{TEC-4}
      & Overall development activity
      & $|\{ \commit \in \Commits \mid \forall \file \in \files(\commit) \cdot \file \not \in \Docs \}|$
      \\ 
    
    \bottomrule
      
  \end{tabular}

\end{table*}

\subsection{Data analysis}\label{subsec:data-analysis}

We follow a Bayesian approach for data analysis. This provides a more nuanced and probabilistic approach compared to traditional frequentist methods \cite{furia2022applying}. Bayesian analysis begins with prior distributions, which summarize plausible parameter values before observing the data. Collected data are included as observations. Then, Bayesian inference is used to redistribute probability over parameter values according to the observations. The outcome is an update of the original prior beliefs, i.e., a posterior distribution over parameter values~\cite{kruschke2018bayesian}. The 95\% High Density Interval (HDI) of the posterior distribution comprises parameter values whose cumulative probability amounts to 95\% of probability density. These parameter values are considered the most credible and viable findings. Parameter values falling in the remaining 5\% cumulative probability are considered non-credible~\cite{kruschke2018bayesian}.

Overall, posterior distributions summarize the relative credibility of all possible parameter values. 
In our analysis, we compute posterior distributions over parameter values that quantify the effect of sustainability metrics on quality metrics.
We quantify this effect for all project types: graduated, retired, graduated post-incubation and bypassed.
Using these posterior distributions, we compare the strength of sustainability on quality for the different project trajectories---i.e., for what project trajectory does sustainability have a stronger effect on quality?.
To this end, we calculate the difference between effect posterior distributions for different project trajectories.
As a result, we obtain posterior distributions that quantify for what trajectory of projects (graduated, retired, post-incubation or bypassed) sustainability metrics have a stronger effect on quality metrics.
Using Bayesian analysis, we are able to not only determine the strength of this difference, but also to quantify its certainty.
To accept or reject our hypotheses, we complement our analysis with a binary criterion.

We define a Bayesian decision criterion based on the HDI of the posterior distributions comparing the strength of sustainability on quality in different project trajectories. 
To determine whether there is a difference on the effect of sustainability on quality, we should find statistical evidence that effect difference between distinct project types is distinct than 0. Hence, the decision criterion is as follows:

\begin{itemize}
    \item [-] If the HDI of the posterior difference includes 0, we conclude that there is no statistical evidence to assert that there exists a difference in the effect of sustainability on quality for the target project trajectories.

    \item [-] If the HDI does not include 0, we conclude that there is statistical evidence to assert that the effect of sustainability on quality is stronger for one of the project trajectories being compared.
      In other words, we have found statistical evidence indicating that an improvement on sustainability leads to a greater improvement in quality for the stronger project trajectory.
      %
    
\end{itemize}
  
The validity of this criterion lies at the definition of HDI. When 0 is included in the HDI, it becomes a plausible value that cannot be discarded; which means that the effect of sustainability on quality is the same for different project trajectories. 
However, when 0 is outside the HDI, it is not considered a credible value---as its associated probability is less than 5\%. 
This, in turn, provides evidence for a stronger effect for one of the project trajectories. 
Due to the units of each metric, the determining what project type has stronger effect is to be interpreted on a case-by-case basis---we describe the details below.

We consider two standard types of Bayesian probabilistic models, depending on the target software quality variable. For \textbf{SWQ-1}, \textbf{SWQ-2.1}, and \textbf{SWQ-2.6}, we use Gaussian regression. As we describe below, we consider the log-transform of these variables. Log-transformed variables are continuous variables with support $(-\infty,+\infty)$. Our Gaussian regression uses a Gaussian distribution as a distribution for the data, which has the required support. Furthermore, the Gaussian distribution is the maximum entropy distribution for a given mean and variance~\cite{mcelreath2020}. This ensures that our Gaussian model does not impose any constraints on the data distribution other than having a fixed mean and finite variance. The remaining software quality variables are model count data. They are discrete variables with support $(0, +\infty)$. The standard data distribution for this type of variable is the Poisson distribution~\cite{mcelreath2020}, as it describes the probability of a given number of events occurring in an interval of time. In our case, these events are the presence of a medium- or high-risk function, a very large file or function, or a line of code in the most complex function. Consequently, we use a Poisson regression model for these variables.

The structure of the model types is as follows (left: Gaussian regression, right: Poisson regression):

\[
\begin{aligned}
\textit{\underline{Gaussian regression}} \hspace{-5cm}&\\
\sigma & \sim \mathcal{U}(10^{-3}, 10) \\
\alpha_i & \sim \mathcal{N}(0, 10) \quad \text{for } i \in \{\text{COM-1}, \ldots, \text{TEC-4}\} \\
\delta_j & \sim \mathcal{N}(0, 1) \quad \text{for } j \in \{\text{non-dormant}, \text{dormant}\} \\
\beta & \sim \mathcal{N}(0, 10) \\
\mu & = \beta + \delta_j + \sum_{i} \alpha_i x_i \\
\log(y) & \sim \mathcal{N}(\mu, \sigma) \\
\end{aligned}
\hspace{3mm}
\begin{aligned}
\textit{\underline{Poisson regression}} \hspace{-4.5cm}&\\
\alpha_i & \sim \mathcal{N}(0, 10) \quad \text{for } i \in \{\text{COM-1}, \ldots, \text{TEC-4}\} \\
\delta_j & \sim \mathcal{N}(0, 1) \quad \text{for } j \in \{\text{non-dormant}, \text{dormant}\} \\
\beta & \sim \mathcal{N}(0, 10) \\
\log(\lambda) & = \beta + \delta_j + \sum_{i} \alpha_i \log(x_i) \\
y & \sim \mathcal{P}(\lambda) \\
\end{aligned}
\]

Our analysis focuses on the effect of each individual sustainability metric on software quality; for each project trajectory. 
Therefore, these model types produce multiple model instances, one for each combination of sustainability metric, quality metric and project trajectory.
We use $y$ to denote the outcome software quality data and $x_i$ the sustainability metric data.
We remark that this analysis differs from computing a correlation coefficient between variables. Correlation coefficients measure the extent to which two variables have a linear relation, e.g., Pearson coefficient (a popular regression coefficient) is a value in $(-1,1)$. Also, correlation coefficients are often symmetric meaning that they do not measure the effect of one variable on the other. They only measure their linear relation. Our Bayesian models are designed to measure the effect of a predicted variable (a sustainability metric) to a predicted variable (a software quality metric). The scale $\alpha_i$ parameter for predictors measures the strength of this effect, i.e., how much a change in sustainability increases quality. This is different from measuring the strength of the linear relation between the variables, which does not have such a direct interpretation. 
Note also the difference in our Bayesian analysis with respect to standard Linear Regression techniques. Standard algorithms for linear regression, such as Least-Squares or Maximum Likelihood Estimation compute the value of coefficients that minimize a mean squared loss or maximize a likelihood function, respectively. These are point estimates, i.e., these methods try to find a single optimal value for each parameter. In the Bayesian setting, we obtain a posterior distribution over parameter values---instead of a single point estimate---which allows us to better understand the uncertainty in the results. This information is of great value, as it helps us draw better-informed conclusions from the analysis.
In summary, our Bayesian analysis provides an easy to interpret and very detailed analysis of the effect of sustainability metrics on quality compared to computing a correlation coefficient. 
In what follows, we describe the specific details of each model type, and how we use them to compare the strength of sustainability on quality for different project types.

\paragraph{Gaussian regression}
We start by performing two data transformations: i) log transform outcome variables and ii) data standardization for predictors. Data standardization consists in performing a linear transformation so that the data have a mean of zero and a standard deviation of one~\cite{kruschke2018bayesian}. This transformation is applied to $x_i$. This process improves the performance of Bayesian inference as the set of possible values for parameters is more concentrated. Computing the log of outcome quality metrics, $\log(y)$, helps to better fit the Gaussian distribution. As a result, we obtain better predictive models for Gaussian regression. The parameter $\alpha_i$ captures whether sustainability metrics have a positive or negative effect on the outcome variable, except for \textbf{STA-6}. Since \textbf{STA-6} is a binary categorical variable, it is modeled as two parameters, $\delta_j$, each of which determines the effect of \textit{non-dormant} and \textit{dormant} projects, respectively. The parameter $\beta$ is the intercept of the model, and $\sigma$ is the standard deviation of the Gaussian data distribution. 
We use a Gaussian, $\mathcal{N}(0,10)$, prior for $\alpha_i, \beta$.
These are non-informative priors that extensively cover all plausible values of the log-transformed outcome variable.
The prior on $\delta_j$ has lower standard deviation, $\mathcal{N}(0,1)$, to prevent undesired interactions with $\beta$ when estimating its value.
We use a uniform prior on positive values, $\mathcal{U}(10^{-3},10)$ for $\sigma$, as we must ensure $\sigma > 0$.
Due to the log transformation of the outcome, model parameters indicate the percentage increase or decrease in the quality metric per unit of the sustainability metric. Since this data is standardized, a unit increase corresponds to one standard deviation increase. Standard deviations for each metric are reported in the accompanying replication package.
 
To study the individual effect of each sustainability metric, we instantiate this model type with $\mu = \beta + \alpha_i x_i$ or $\mu = \delta_j$, depending on the sustainability metric under analysis. Since this model applies to 3 software quality metrics, all sustainability metrics and the four project types, we study $3 \cdot 16 \cdot 4 = 192$ models of this type.

\paragraph{Poisson regression}

This type of model uses a Poisson distribution for software quality metrics, which is commonly used for count variables~\cite{mcelreath2020}. Consequently, we do not standardize the quality metrics that this model analyzes. Standardizing these metrics would transform them into continuous variables, and consequently, we would lose the information about the counting process that produces values for these software quality metrics.

Instead, we perform a log transformation of the data for sustainability metrics, $\log(x_i)$, and keep quality metrics, $y$, unmodified. The log transformation allows us to concentrate the range of possible values for the sustainability parameters $\alpha_i$. However, due to the large number of negative values and zeros in \textbf{STA-4}, \textbf{STA-7}, and \textbf{TEC-1}, we discard them for this type of model; recall log transforms are undefined for these values. The meaning of the parameters $\alpha_i$, $\delta_j$, and $\beta$ is the same as before. 
As priors, we use Gaussian distributions centered at 0, with a standard deviation of 10 (for $\alpha_i$ and $\beta$) and 1 (for $\delta_j$). As before, these are non-informative priors. Note that, as usual for Poisson regression, we use a logarithmic link function for the rate of the Poisson data distribution $\log(\lambda)$~\cite{kruschke2018bayesian}. Thus, our prior on the intercept $\beta$ covers a range $e^{-20} \approx 0$ to $e^{20} \approx 4.85 \times 10^8$. This range captures all plausible values for the target quality metrics and ensures that they are assigned non-zero prior probability. The log transformation of predictors implies that a $10\%$ increase in sustainability changes  $0.1\alpha_i$ or $0.1\delta_j$ units in the code quality metric.

As before, Poisson regression models are instantiated for single sustainability metrics as $\log(\lambda) = \beta + \delta_j$ for \textbf{STA-6} and $\log(\lambda) = \beta + \alpha_i x_i$ for the rest. This model applies to 5 software quality metrics, 13 sustainability metrics and 4 project types; hence, we study $5 \cdot 13 \cdot 4 = 260$ models of this type.

\paragraph{Effect difference for different project trajectories}
To compare the difference on the effect of sustainability on quality, we compute the difference of scale parameters for different project trajectories (graduated, retired, post-incubation, bypassed).
As mentioned above, the parameter $\alpha$ quantifies the effect of sustainability on quality and, for \textbf{STA-6} it is $\delta$---for brevity in what follows we refer only to $\alpha$ but the same operations are computed on $\delta$ for the cases regarding \textbf{STA-6}.
Let $t,u$ denote two project types and $\alpha^t$, $\alpha^u$ denote the effect for the same sustainability and quality metrics but for the different project types.
Then, the effect difference is computed as $\Delta = \alpha^t - \alpha^u$.
Determining what project type has stronger effect depends on the sustainability/quality metrics under analysis.
If an increment in the sustainability/quality metrics captures an improvement in sustainability/quality, respectively, then positive values of $\Delta$ indicate a stronger effect of $t$ with respect to $u$.
As an example of this situation, consider sustainability metric \textbf{COM-2} (frequency of communication) and quality metric \textbf{SWQ-2.1} (code coverage).
An increment in the frequency of communication implies an improvement in sustainability, and, similarly, an increment in code coverage results in a improvement in software quality.
If an increment in the sustainability metric improves sustainability and a decrement of the quality metric improves quality (or vice versa), then negative values of $\Delta$ indicate a stronger effect of $t$ with respect to $u$.
This occurs for instance, when studying \textbf{COM-2} (frequency of communication)/\textbf{SWQ-1} (defect density) or \textbf{COM-1} (response time)/\textbf{SWQ-2.1} (code coverage).
Finally, if a decrement in the sustainability/quality metrics captures an improvement in sustainability/quality, respectively, then negative values of $\Delta$ indicate a stronger effect of $t$ with respect to $u$.
This is the case, for instance, when analyzing \textbf{COM-1}/\textbf{SWQ-1}.
The following section describes this analysis in the context of hypothesis \textbf{H1} (see \ref{sec:H1}).

To perform Bayesian inference, we implemented our models in the probabilistic programming library PyMC~\cite{pymc}. Since all the model parameters are continuous, we use the NUTS sampler, which is the best performing sampler for estimating continuous parameters~\cite{nuts}. For each model, we computed four parallel chains of 3000 samples each. All our analyses show a Monte Carlo Standard Error (MCSE) below 0.02, which indicates high accuracy~\cite{kruschke2018bayesian}. Furthermore, we also perform a posterior predictive check, which shows that our models have good predictive accuracy.

%% file: findings.tex
\section{Findings}\label{sec:findings}


In this study, we sought to investigate how project trajectories (i.e., \textbf{Graduated,} \textbf{Post-Graduation,} \textbf{Retired,} and \textbf{Bypassed}) moderate the relationship between sustainability and software quality. We analyzed how the sustainability-quality relationship manifests across graduated, post-graduation, retired, or bypassed incubation projects. Our findings show that sustainability metrics are more strongly associated with software quality in graduated and bypassed projects than in retired ones. For graduated projects, this relationship shows an overall strengthening post-incubation. Accordingly, all our hypotheses are supported, except \textbf{H3.} We found that graduated projects that progress through incubation exhibit a stronger sustainability-quality relationship compared to bypassed, opposed to our expectations, underscoring the importance of structured governance and adherence to sustainability practices in shaping software quality outcomes. In the following subsections, we provide a detailed breakdown of the findings and the analysis that validate our hypotheses.

\begin{figure}[th!]

  \centering

  \input{tables_automatically_generated/table_h1.tex}

  \caption{\textbf{H1 Comparison of Sustainability Metrics' Impact on SWQ Across Graduated and Retired Projects During Incubation.}  Cells labeled \textbf{``Graduated''} indicate that the given sustainability metric has a stronger effect on SWQ in graduated projects, whereas \textbf{``Retired''} signifies a stronger effect in retired projects. Cells marked \textbf{\ding{56}} indicate that there is no significant difference on the impact of the sustainability metric on SWQ in the different project trajectories. \textbf{``NA''} represents cases where computational limitations prevented evaluation.}%
  
  \label{fig:H1_table}
  
\end{figure}

\subsection{H1: Graduated vs. Retired During Incubation}\label{sec:H1}

We hypothesized that \emph{graduated projects exhibit a stronger positive relationship between sustainability and software quality metrics compared to retired projects during the incubation period}, to evaluate the moderating effect of incubation as a development trajectory on the sustainability-quality relationship of graduated and retired projects. Our analysis confirms this hypothesis, indicating that sustainability metrics have a stronger positive effect on software quality in graduated projects compared to the retired ones.

Figure~\ref{fig:H1_table} summarizes our comparative analysis, highlighting that the sustainability-quality relationship is stronger in 35 instances for graduated compared to 15 for retired project. For example, \textbf{STA-4} (\emph{growth}) and \textbf{STA-7} (\emph{retention}) have stronger effect on \textbf{SWQ-1} (\emph{defect density}) in graduated projects compared to retired projects, whereas, \textbf{COM-2} (\emph{frequency of communication}) has stronger effect on \textbf{SWQ-1} (\emph{Code Duplication Percentage}) for retired projects compared to graduated projects. Overall, we conclude that our selected sustainability metrics exert a stronger positive influence on software quality in graduated projects.

We reached the outcome summarized in Fig.~\ref{fig:H1_table} after conducting an extensive Bayesian statistical analysis to evaluate the effect of our selected sustainability metrics on software quality. For the sake of brevity and to ensure a concise presentation, we will use \textbf{STA-4} (\emph{growth}) effect on \textbf{SWQ-1} (\emph{defect density}), \textbf{COM-2} (\emph{frequency of communication}) effect on \textbf{SWQ-2.1} (\emph{code coverage}), and \textbf{STA-9} (\emph{turnover}) effect on \textbf{SWQ-2.5} (\emph{Very Large Function Size Count}) to describe the details of the analysis. We conducted similar analysis for the other metrics. Similar analysis was carried out for every sustainability and SWQ metrics combination for all hypotheses. The outcome of this detailed analysis can be found in our replication package (see Sect. \ref{sec:replication}).
                
\begin{figure}[th!]

  \centering
    \includegraphics[ trim=0cm 0cm 0cm 14mm, clip,
     width = 1.0 \textwidth
    ]{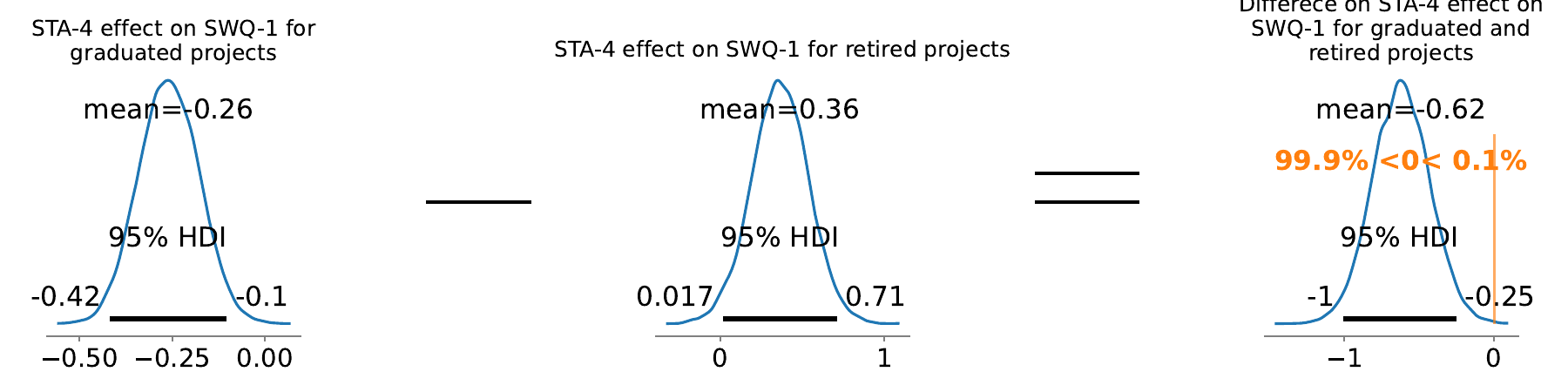}

    \caption{Posterior plots for the impact of \emph{growth} (\textbf{STA-4}) on defect density (\textbf{SWQ-1}) for graduated (left) and retired projects (middle), and the posterior effect difference (right).}%
    
    \label{fig:H1_plot1}
    
\end{figure}

To further illustrate the analysis underpinning the result of \textbf{H1}, Fig. \ref{fig:H1_plot1} presents the posterior distribution on the difference of the effect of \textbf{STA-4} (\emph{growth}) on \textbf{SWQ-1} (\emph{defect density}), comparing graduated and retired projects (right plot). 

The left and middle posterior distributions correspond to the effect of \textbf{STA-4} on \textbf{SWQ-1} for graduated and retired projects, respectively. We observe a negative value for graduated projects, which implies that incremental growth decreases defect density. For retired projects, we primarily observe positive values, indicating an increase in defect density despite experiencing growth.

After completing the analysis for both retired and graduated projects in isolation, we conducted a comparative analysis. We compare the effects on these two groups of projects by computing the effect difference (right plot). The plot shows that the mean effect difference is \textbf{-0.62}, which means that higher growth rates in a project are linked to a greater drop in defect density for graduated projects. The 95\% Highest Density Interval (HDI), $(-1, -0.25)$, excludes zero, which confirms statistically the higher reduction for graduated projects. Therefore, we conclude that the growth effect on reducing defect density in graduated projects is stronger.

This finding reinforces the argument that a steady influx of contributions enhances defect management, as observed in a reduction in defect density in graduated projects. In contrast, retired projects exhibit weaker effects, suggesting that even with supported structured mentoring, and governance, retired projects during incubation experience negative effects on defect density, when experiencing growth, an important aspect of SWQ.

This analysis highlights that while experiencing growth (increased PRs and feature development), graduated projects during incubation maintain consistent defect resolution. However, the retired sample does not exhibit similar behavior. This could be explained by the motivation of the graduated project to graduate successfully when maintaining low defect counts.

\begin{figure}[th!]

  \centering
    \includegraphics[ trim=0cm 0cm 0cm 1.2cm, clip,
     width = 0.8 \textwidth
    ]{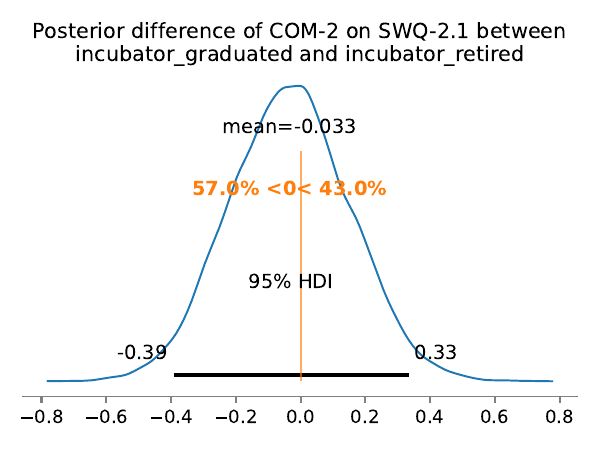}

    \caption{Posterior plots for the effect difference of \emph{frequency of communication} (\textbf{COM-2}) on code coverage (\textbf{SWQ-2.1}) for graduated and retired projects.}%
    
    \label{fig:H1_plot2}
    
\end{figure}

To further illustrate our analysis, Fig. \ref{fig:H1_plot2} presents the posterior distribution on the difference of the effect of \textbf{COM-2} (\emph{frequency of communication}) on \textbf{SWQ-2.1} (\emph{code coverage}), comparing graduated and retired projects. 
The posterior plot indicates a mean effect of \textbf{-0.033}, suggesting a minor stronger positive effect of increased communication frequency on code coverage for graduated projects. However, the 95\% HDI ranges from $-0.39$ to $0.33$, including zero, indicates that there is no statistical evidence to assert this effect.

This suggests that communication frequency, as a sustainability indicator, effect on code coverage maybe dependent on other interacting factors such as the nature of discussions, the quality of the communication, and the timing of the communication. Unlike growth, which exhibited a stronger effect on defect density, the communication frequency's impact on code coverage could be more complex and context-dependent.

\begin{figure}[th!]

  \centering
    \includegraphics[ trim=0cm 0cm 0cm 1.2cm, clip,
     width = 0.8 \textwidth
    ]{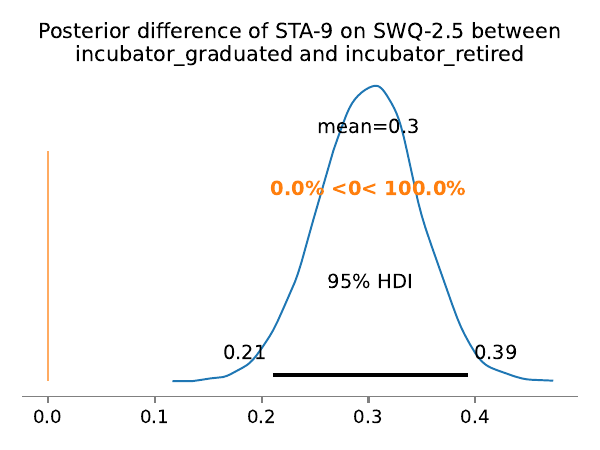}

    \caption{Posterior plots for the effect difference of \emph{turnover} (\textbf{STA-9}) on \emph{Very Large Function Size Count} (\textbf{SWQ-2.5}) for graduated and retired projects.}%
    
    \label{fig:H1_plot3}
    
\end{figure}

Similarly, we examined the effect of \textbf{STA-9} (\emph{turnover}) on \textbf{SWQ-2.5} (\emph{Large Function Size Count}), comparing graduated and retired projects. While an increase in turnover shows resilience among graduated projects, i.e., a stronger effect on code quality metrics despite a reduction in the population of contributors, the effect on \textbf{SWQ-2.5} was particularly stronger for retired projects. This is highly likely an isolated instance rather than a recurrent behavior, given it is inconsistent for other quality metrics.

Figure~\ref{fig:H1_plot3} presents the posterior distribution of the effect of \textbf{STA-9} (\emph{turnover}) on \textbf{SWQ-2.5} (\emph{Very Large Function Size Count}), comparing graduated and retired projects. The posterior plot indicates a mean effect difference of \textbf{0.3}, and the 95\% HDI ranges from $0.21$ to $0.39$, excluding zero, confirming statistically a stronger positive effect for retired projects.

While turnover implies a loss of contributors, their expertise and knowledge, its effect in graduated projects appears to be mitigated (stronger effect on four code quality metrics, see \ref{fig:H1_table}), highly likely with established development practices and the motivation to graduate. This is reflected in a more stable quality metrics despite fluctuations in contributor population. In retired projects, while we observed stronger resilience in one particular metric, i.e., very large function sizes, Fig. \ref{fig:H1_table} suggests a lack of consistency in coding practices. For example, the stronger effect observed on \textbf{SWQ-2.5} is not mirrored on other metrics, such as \textbf{SWQ-2.4} (\emph{number of very large files }) and \textbf{SWQ-2.7} (\emph{Most Complex Function LOC}) opposed to the graduated projects.


\subsection{H2: After Graduation}

We hypothesized that \emph{graduated projects exhibit a stronger positive relationship between sustainability and software quality metrics post-graduation compared to the incubation period}, to evaluate the moderating effect of the post-incubation period as a development trajectory on the sustainability-quality relationship of graduated projects. Our analysis support this hypothesis. Figure \ref{fig:H2_table} shows that sustainability metrics have a stronger positive effect on software quality metrics post incubation. 

\begin{figure}[th!]

  \centering

  \input{tables_automatically_generated/table_h2.tex}

  \caption{\textbf{H2 Comparison of Sustainability Metrics' Impact on SWQ Across Graduated Projects During Incubation and Post Incubation.}  Cells labeled \textbf{``Incubation''} indicate that the given sustainability metric has a stronger effect on SWQ during incubation, whereas \textbf{``Post incubation''} signifies a stronger effect post incubation (or after graduation). Cells marked \textbf{\ding{56}} indicate that there is no significant difference on the impact of the sustainability metric on SWQ in the compared periods. \textbf{``NA''} represents cases where computational limitations prevented evaluation.}%
  
  \label{fig:H2_table}
  
\end{figure}

To illustrate this conclusion, we will use \textbf{COM-2} (response time) effect on \textbf{SWQ-1} (\emph{defect density}) and \textbf{STA-4} (growth) effect on \textbf{SWQ-2.6} (\emph{very large function size count}) analysis.

\begin{figure}[th!]

  \centering
    \includegraphics[ trim=0cm 0cm 0cm 1.2cm, clip,
     width = 0.8 \textwidth
    ]{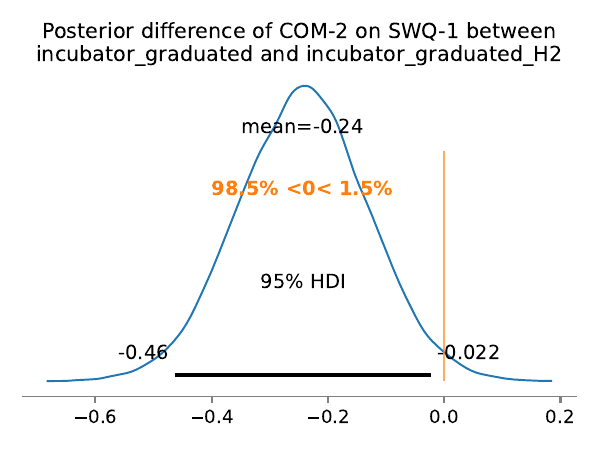}

    \caption{Posterior plots for the effect difference of \emph{response time} (\textbf{COM-2}) on \emph{defect density} (\textbf{SWQ-1}) for incubating and post-incubation projects.}%
    
    \label{fig:H2_plot1}
    
\end{figure}

Figure \ref{fig:H2_plot1} presents the posterior distribution of the effect of \textbf{COM-2} (\emph{response time}) on \textbf{SWQ-1} (\emph{defect density}), comparing graduated projects during incubation and post-incubation. It shows a mean effect of \textbf{-0.24}, suggesting that a decrease in response time (i.e., quicker responses) is associated with a stronger reduction in defect density for incubating projects. The 95\% HDI ranges from $-0.46$ to $-0.022$, excluding zero, which confirms statistically the stronger effect during incubation.

This finding suggests that before graduation, projects seem to optimize communication efficiency, reducing response times to comments, in issue tracking, and code reviews. This immediate feedback loop appears to enhance defect resolution. Post-incubation, the priority shifts from defect resolution to the structural aspect of code, as shown by the increase effect of sustainability metrics on code quality attributes like code complexity and duplication.

Overall, and as shown in Fig.~\ref{fig:H2_table}, 20 sustainability metrics exert a stronger influence on software quality post-incubation, compared to 12 during incubation. This suggests a shift in focus from defect reduction during incubation to broader structural quality improvements after graduation. Reduced defects count is a key driver for graduation, aligning with ASF's expectations. This could explain the behavior we observed in this analysis. The shift to the structural aspects of code quality, could be also explained by the alleviation of the graduation burden. Once graduated, projects may no longer prioritize meeting incubation requirements, such as rigorous defect resolution, to prove their viability. Subsequently, they seem to shift their focus toward long-term code quality and maintainability.

\begin{figure}[th!]

  \centering
    \includegraphics[ trim=0cm 0cm 0cm 1.2cm, clip,
     width = 0.8 \textwidth
    ]{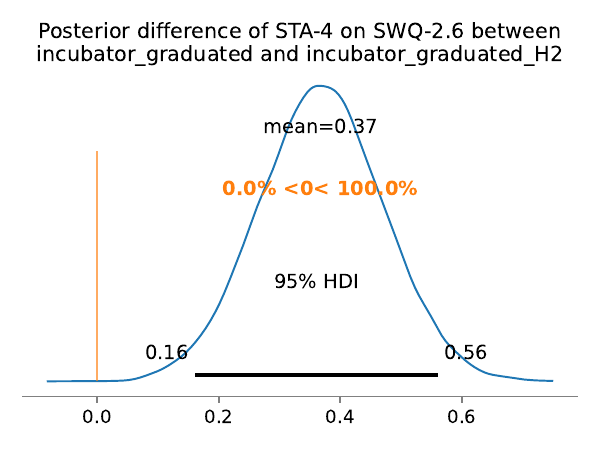}

    \caption{Posterior plots for the effect difference of \emph{growth} (\textbf{STA-4}) on \emph{code duplication percentage} (\textbf{SWQ-2.6}) for incubating and post-incubation projects.}%
    
    \label{fig:H2_plot2}
    
\end{figure}

Figure \ref{fig:H2_plot2} presents the posterior distribution of the effect of \textbf{STA-4} (\emph{growth}) on \textbf{SWQ-2.6} (\emph{code duplication percentage}), comparing graduated projects during incubation and post-incubation. The mean effect is \textbf{0.37}, with a 95\% HDI ranging from $0.16$ to $0.56$, indicating a statistically credible positive effect post-incubation.

This result suggests that post-incubation, despite increase in PRs and an evolving codebase, projects maintain consistent coding practices, as observed in stronger effects on \textbf{SWQ-2.1}-\textbf{SWQ-2.7}. Unlike incubation, where rapid resolution of defects may take precedence, post-incubation, projects seem to focus on long-term maintainability. This aligns with our interpretation that while projects experience growth in feature development after incubation, they seem to prioritize stability by shifting their code quality practices from immediate defect reduction to broader structural concerns.

\subsection{H3: Bypassed vs. Graduated}

\begin{figure}[th!]

  \input{tables_automatically_generated/table_h3.tex}

  \caption{\textbf{H3 Comparison of Sustainability Metrics' Impact on SWQ Across Graduated and Bypassed Projects.}  Cells labeled \textbf{``Graduated''} indicate that the given sustainability metric has a stronger effect on SWQ for graduated projects, whereas \textbf{``Bypassed''} signifies a stronger effect for bypassed projects. Cells marked \textbf{\ding{56}} indicate that there is no significant difference on the impact of the sustainability metric on SWQ in the compared populations, graduated and bypassed. \textbf{``NA''} represents cases where computational limitations prevented evaluation.}%
  
  \label{fig:H3_table}
  
\end{figure}

It is suggested in \textbf{H3} that \emph{bypassed projects exhibit an equal or stronger positive relationship between sustainability and software quality metrics compared to graduated projects post-incubation.} We based this hypothesis on the premise that bypassed projects have developed stronger and more resilient alternative mechanisms for sustainability due to the lack of structured governance and mentorship of the incubation process. In addition, these projects may have stronger internal resilience because they are older, matured, and attracted highly committed contributors over time. We expected that this resilience would translate into stronger quality practices compared to those that graduated through the formal incubation process. However, our hypothesis (\textbf{H3}) does not hold.

\begin{figure}[th!]

  \centering
    \includegraphics[ trim=0cm 0cm 0cm 1.2cm, clip,
     width = 0.8 \textwidth
    ]{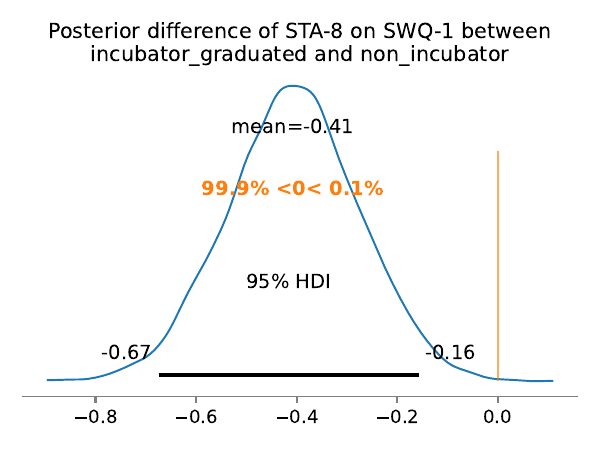}

    \caption{Posterior plots for the effect difference of \emph{size} (\textbf{STA-8}) on \emph{defect density} (\textbf{SWQ-1}) for graduated and bypassed projects.}%
    
    \label{fig:H3_plot1}
    
\end{figure}

Figure~\ref{fig:H3_table} shows that sustainability metrics have a stronger positive effect on software quality metrics for graduated projects compared to bypassed. Graduated project exhibited stronger effects in 35 instances compared to 22 for bypassed projects.

We will use \textbf{STA-8} (size) effect on \textbf{SWQ-1} (\emph{defect density}) and \textbf{COM-2} (frequency of communication) effect on \textbf{SWQ-2.7} (\emph{most complex function LOC}) as illustrative cases.

Figure \ref{fig:H3_plot1} presents the posterior distribution of the effect of \textbf{STA-8} (\emph{size}) on \textbf{SWQ-1} (\emph{defect density}), comparing graduated and bypassed projects. The mean effect is \textbf{-0.41}, with a 95\% HDI ranging from $-0.67$ to $-0.16$, statistically indicating a stronger reduction on defect density among graduated projects. The posterior probability of this effect is 99.9\%. This suggests that the structured governance and formal incubation process in graduated projects contribute to long term adherence to sustainability practices and improved defect management, whereas bypassed projects exhibit weaker effects in this regard.

\begin{figure}[th!]

  \centering
    \includegraphics[ trim=0cm 0cm 0cm 1.2cm, clip,
     width = 0.8 \textwidth
    ]{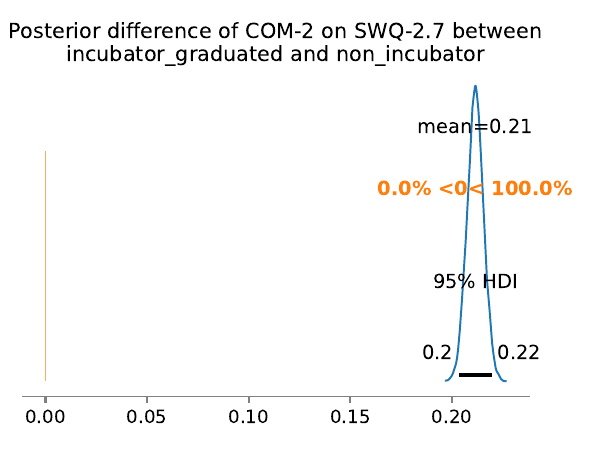}

    \caption{Posterior plots for the effect difference of \emph{communication frequency} (\textbf{COM-2}) on \emph{most complex function LOC} (\textbf{SWQ-2.7}) for graduated and bypassed projects.}%
    
    \label{fig:H3_plot2}
    
\end{figure}

Figure \ref{fig:H3_plot2} presents the posterior distribution of the effect of \textbf{COM-2} (\emph{communication frequency}) on \textbf{SWQ-2.7} (\emph{most complex function LOC}), comparing graduated and bypassed projects. The mean effect is \textbf{0.29}, with a 95\% HDI ranging from $0.08$ to $0.51$, indicating a statistically credible stronger effect among bypassed projects. The posterior probability that the effect is stronger for bypassed projects is 99.4\%, suggesting that increased communication frequency is associated with lower complexity in function size among bypassed projects. This trend is also observed across other sustainability metrics affecting \textbf{SWQ-2.7}. A possible explanation is that bypassed projects are probably more attentive to the complexity of their code. This particular attention could be explained by their sensitivity to complexity as their codebase grows, especially because they are older than the graduated ones.

We had a similar observation in our previous work \cite{alami2024free}, both retired and graduated projects combined in a single sample, exhibited improvements in metrics related to code complexity (\textbf{SWQ-2.3 ... SWQ-2.7}), when project's age increases. However, that is not the case for defect density (\textbf{SWQ-1}). These findings may suggest that older projects become more mature at prioritizing and effectively managing code quality, addressing issues of complexity and maintainability \cite{alami2024free}. 

\subsection{H4: Bypassed vs. Graduated vs. Retired}

In \textbf{H4}, we proposed that \emph{the relationship between sustainability metrics and software quality is moderated by project trajectory, with bypassed and graduated projects showing a stronger relationship than retired projects.} This hypothesis holds. Figures~\ref{fig:H1_table} and \ref{fig:H4_table} show that sustainability metrics have stronger effects on SWQ in 35 instances for graduated compared to only 15 for retired. Similarly, when comparing bypassed and retired, the effect was stronger 34 times compared to 20.

\begin{figure}[th!]

  \input{tables_automatically_generated/table_h4_bypassed_retired.tex}

  \caption{\textbf{H4 Comparison of Sustainability Metrics' Impact on SWQ Across Retired and Bypassed Projects.}  Cells labeled \textbf{``Retired''} indicate that the given sustainability metric has a stronger effect on SWQ for retired projects, whereas \textbf{``Bypassed''} signifies a stronger effect for bypassed projects. Cells marked \textbf{\ding{56}} indicate that there is no significant difference on the impact of the sustainability metric on SWQ in the compared populations, retired or bypassed. \textbf{``NA''} represents cases where computational limitations prevented evaluation.}%
  
  \label{fig:H4_table}
  
\end{figure}

In brief, these findings demonstrate that project trajectories have a role in shaping the sustainability-SWQ relationship. We found that graduated projects benefit the most from sustainability practices, exhibiting stronger code quality, both during and after incubation. Post-incubation, the influence of sustainability metrics on SWQ becomes even more pronounced, suggesting a shift from immediate defect resolution toward long-term maintainability. Contrary to our expectations, bypassed projects did not outperform graduated projects. Instead, we found that structured governance and mentoring during incubation appeared to provide a more consistent pathway to sustainability and enhanced software quality practices. However, bypassed projects showed a unique sensitivity to code complexity, likely reflecting their self-regulated development strategies and the experience they gained from an evolving codebase over time. Finally, retired projects exhibited the weakest sustainability-SWQ relationship, supporting our original claim that weaker sustainability practices are associated with diminished software quality outcomes.

%% file: tables_automatically_generated/table_h1.tex
\bgroup
\def\arraystretch{1.4}
\scalebox{0.9}{
\begin{tabularx}{1.1\textwidth}{
  p{25mm}
  >{\centering\arraybackslash}m{14mm}
  >{\centering\arraybackslash}m{14mm}
  >{\centering\arraybackslash}m{14mm}
  >{\centering\arraybackslash}m{14mm}
  >{\centering\arraybackslash}m{14mm}
  >{\centering\arraybackslash}m{14mm}
  >{\centering\arraybackslash}m{14mm}
  >{\centering\arraybackslash}m{14mm}
}
\toprule
 & \textbf{SWQ-1} & \textbf{SWQ-2.1} & \textbf{SWQ-2.6} & \textbf{SWQ-2.2} & \textbf{SWQ-2.3} & \textbf{SWQ-2.4} & \textbf{SWQ-2.5} & \textbf{SWQ-2.7} \\
\midrule
\textbf{COM-1} & {\cellcolor[HTML]{40B5C4}} \color[HTML]{F1F1F1} \ding{56} & {\cellcolor[HTML]{40B5C4}} \color[HTML]{F1F1F1} \ding{56} & {\cellcolor[HTML]{24419A}} \color[HTML]{F1F1F1} Graduated & {\cellcolor[HTML]{24419A}} \color[HTML]{F1F1F1} Graduated & {\cellcolor[HTML]{40B5C4}} \color[HTML]{F1F1F1} \ding{56} & {\cellcolor[HTML]{40B5C4}} \color[HTML]{F1F1F1} \ding{56} & {\cellcolor[HTML]{40B5C4}} \color[HTML]{F1F1F1} \ding{56} & {\cellcolor[HTML]{E1F3B2}} \color[HTML]{000000} Retired \\
\textbf{COM-2} & {\cellcolor[HTML]{40B5C4}} \color[HTML]{F1F1F1} \ding{56} & {\cellcolor[HTML]{40B5C4}} \color[HTML]{F1F1F1} \ding{56} & {\cellcolor[HTML]{E1F3B2}} \color[HTML]{000000} Retired & {\cellcolor[HTML]{24419A}} \color[HTML]{F1F1F1} Graduated & {\cellcolor[HTML]{E1F3B2}} \color[HTML]{000000} Retired & {\cellcolor[HTML]{40B5C4}} \color[HTML]{F1F1F1} \ding{56} & {\cellcolor[HTML]{E1F3B2}} \color[HTML]{000000} Retired & {\cellcolor[HTML]{E1F3B2}} \color[HTML]{000000} Retired \\
\textbf{POP-1} & {\cellcolor[HTML]{40B5C4}} \color[HTML]{F1F1F1} \ding{56} & {\cellcolor[HTML]{24419A}} \color[HTML]{F1F1F1} Graduated & {\cellcolor[HTML]{E1F3B2}} \color[HTML]{000000} Retired & {\cellcolor[HTML]{24419A}} \color[HTML]{F1F1F1} Graduated & {\cellcolor[HTML]{24419A}} \color[HTML]{F1F1F1} Graduated & {\cellcolor[HTML]{24419A}} \color[HTML]{F1F1F1} Graduated & {\cellcolor[HTML]{40B5C4}} \color[HTML]{F1F1F1} \ding{56} & {\cellcolor[HTML]{24419A}} \color[HTML]{F1F1F1} Graduated \\
\textbf{STA-1} & {\cellcolor[HTML]{40B5C4}} \color[HTML]{F1F1F1} \ding{56} & {\cellcolor[HTML]{40B5C4}} \color[HTML]{F1F1F1} \ding{56} & {\cellcolor[HTML]{40B5C4}} \color[HTML]{F1F1F1} \ding{56} & {\cellcolor[HTML]{24419A}} \color[HTML]{F1F1F1} Graduated & {\cellcolor[HTML]{40B5C4}} \color[HTML]{F1F1F1} \ding{56} & {\cellcolor[HTML]{40B5C4}} \color[HTML]{F1F1F1} \ding{56} & {\cellcolor[HTML]{E1F3B2}} \color[HTML]{000000} Retired & {\cellcolor[HTML]{40B5C4}} \color[HTML]{F1F1F1} \ding{56} \\
\textbf{STA-2} & {\cellcolor[HTML]{40B5C4}} \color[HTML]{F1F1F1} \ding{56} & {\cellcolor[HTML]{40B5C4}} \color[HTML]{F1F1F1} \ding{56} & {\cellcolor[HTML]{40B5C4}} \color[HTML]{F1F1F1} \ding{56} & {\cellcolor[HTML]{40B5C4}} \color[HTML]{F1F1F1} \ding{56} & {\cellcolor[HTML]{24419A}} \color[HTML]{F1F1F1} Graduated & {\cellcolor[HTML]{24419A}} \color[HTML]{F1F1F1} Graduated & {\cellcolor[HTML]{24419A}} \color[HTML]{F1F1F1} Graduated & {\cellcolor[HTML]{24419A}} \color[HTML]{F1F1F1} Graduated \\
\textbf{STA-3} & {\cellcolor[HTML]{40B5C4}} \color[HTML]{F1F1F1} \ding{56} & {\cellcolor[HTML]{24419A}} \color[HTML]{F1F1F1} Graduated & {\cellcolor[HTML]{40B5C4}} \color[HTML]{F1F1F1} \ding{56} & {\cellcolor[HTML]{24419A}} \color[HTML]{F1F1F1} Graduated & {\cellcolor[HTML]{24419A}} \color[HTML]{F1F1F1} Graduated & {\cellcolor[HTML]{24419A}} \color[HTML]{F1F1F1} Graduated & {\cellcolor[HTML]{40B5C4}} \color[HTML]{F1F1F1} \ding{56} & {\cellcolor[HTML]{24419A}} \color[HTML]{F1F1F1} Graduated \\
\textbf{STA-4} & {\cellcolor[HTML]{24419A}} \color[HTML]{F1F1F1} Graduated & {\cellcolor[HTML]{40B5C4}} \color[HTML]{F1F1F1} \ding{56} & {\cellcolor[HTML]{40B5C4}} \color[HTML]{F1F1F1} \ding{56} & {\cellcolor[HTML]{000000}} \color[HTML]{F1F1F1} NA & {\cellcolor[HTML]{000000}} \color[HTML]{F1F1F1} NA & {\cellcolor[HTML]{000000}} \color[HTML]{F1F1F1} NA & {\cellcolor[HTML]{000000}} \color[HTML]{F1F1F1} NA & {\cellcolor[HTML]{000000}} \color[HTML]{F1F1F1} NA \\
\textbf{STA-5} & {\cellcolor[HTML]{40B5C4}} \color[HTML]{F1F1F1} \ding{56} & {\cellcolor[HTML]{40B5C4}} \color[HTML]{F1F1F1} \ding{56} & {\cellcolor[HTML]{40B5C4}} \color[HTML]{F1F1F1} \ding{56} & {\cellcolor[HTML]{24419A}} \color[HTML]{F1F1F1} Graduated & {\cellcolor[HTML]{40B5C4}} \color[HTML]{F1F1F1} \ding{56} & {\cellcolor[HTML]{40B5C4}} \color[HTML]{F1F1F1} \ding{56} & {\cellcolor[HTML]{E1F3B2}} \color[HTML]{000000} Retired & {\cellcolor[HTML]{24419A}} \color[HTML]{F1F1F1} Graduated \\
\textbf{STA-6} {\tiny (dormant)} & {\cellcolor[HTML]{40B5C4}} \color[HTML]{F1F1F1} \ding{56} & {\cellcolor[HTML]{40B5C4}} \color[HTML]{F1F1F1} \ding{56} & {\cellcolor[HTML]{40B5C4}} \color[HTML]{F1F1F1} \ding{56} & {\cellcolor[HTML]{40B5C4}} \color[HTML]{F1F1F1} \ding{56} & {\cellcolor[HTML]{40B5C4}} \color[HTML]{F1F1F1} \ding{56} & {\cellcolor[HTML]{40B5C4}} \color[HTML]{F1F1F1} \ding{56} & {\cellcolor[HTML]{40B5C4}} \color[HTML]{F1F1F1} \ding{56} & {\cellcolor[HTML]{40B5C4}} \color[HTML]{F1F1F1} \ding{56} \\
\textbf{STA-6} {\tiny (non dormant)} & {\cellcolor[HTML]{40B5C4}} \color[HTML]{F1F1F1} \ding{56} & {\cellcolor[HTML]{40B5C4}} \color[HTML]{F1F1F1} \ding{56} & {\cellcolor[HTML]{40B5C4}} \color[HTML]{F1F1F1} \ding{56} & {\cellcolor[HTML]{40B5C4}} \color[HTML]{F1F1F1} \ding{56} & {\cellcolor[HTML]{40B5C4}} \color[HTML]{F1F1F1} \ding{56} & {\cellcolor[HTML]{40B5C4}} \color[HTML]{F1F1F1} \ding{56} & {\cellcolor[HTML]{40B5C4}} \color[HTML]{F1F1F1} \ding{56} & {\cellcolor[HTML]{40B5C4}} \color[HTML]{F1F1F1} \ding{56} \\
\textbf{STA-7} & {\cellcolor[HTML]{24419A}} \color[HTML]{F1F1F1} Graduated & {\cellcolor[HTML]{40B5C4}} \color[HTML]{F1F1F1} \ding{56} & {\cellcolor[HTML]{40B5C4}} \color[HTML]{F1F1F1} \ding{56} & {\cellcolor[HTML]{000000}} \color[HTML]{F1F1F1} NA & {\cellcolor[HTML]{000000}} \color[HTML]{F1F1F1} NA & {\cellcolor[HTML]{000000}} \color[HTML]{F1F1F1} NA & {\cellcolor[HTML]{000000}} \color[HTML]{F1F1F1} NA & {\cellcolor[HTML]{000000}} \color[HTML]{F1F1F1} NA \\
\textbf{STA-8} & {\cellcolor[HTML]{40B5C4}} \color[HTML]{F1F1F1} \ding{56} & {\cellcolor[HTML]{40B5C4}} \color[HTML]{F1F1F1} \ding{56} & {\cellcolor[HTML]{40B5C4}} \color[HTML]{F1F1F1} \ding{56} & {\cellcolor[HTML]{24419A}} \color[HTML]{F1F1F1} Graduated & {\cellcolor[HTML]{24419A}} \color[HTML]{F1F1F1} Graduated & {\cellcolor[HTML]{24419A}} \color[HTML]{F1F1F1} Graduated & {\cellcolor[HTML]{E1F3B2}} \color[HTML]{000000} Retired & {\cellcolor[HTML]{24419A}} \color[HTML]{F1F1F1} Graduated \\
\textbf{STA-9} & {\cellcolor[HTML]{40B5C4}} \color[HTML]{F1F1F1} \ding{56} & {\cellcolor[HTML]{40B5C4}} \color[HTML]{F1F1F1} \ding{56} & {\cellcolor[HTML]{40B5C4}} \color[HTML]{F1F1F1} \ding{56} & {\cellcolor[HTML]{24419A}} \color[HTML]{F1F1F1} Graduated & {\cellcolor[HTML]{24419A}} \color[HTML]{F1F1F1} Graduated & {\cellcolor[HTML]{24419A}} \color[HTML]{F1F1F1} Graduated & {\cellcolor[HTML]{E1F3B2}} \color[HTML]{000000} Retired & {\cellcolor[HTML]{24419A}} \color[HTML]{F1F1F1} Graduated \\
\textbf{TEC-1} & {\cellcolor[HTML]{40B5C4}} \color[HTML]{F1F1F1} \ding{56} & {\cellcolor[HTML]{40B5C4}} \color[HTML]{F1F1F1} \ding{56} & {\cellcolor[HTML]{40B5C4}} \color[HTML]{F1F1F1} \ding{56} & {\cellcolor[HTML]{000000}} \color[HTML]{F1F1F1} NA & {\cellcolor[HTML]{000000}} \color[HTML]{F1F1F1} NA & {\cellcolor[HTML]{000000}} \color[HTML]{F1F1F1} NA & {\cellcolor[HTML]{000000}} \color[HTML]{F1F1F1} NA & {\cellcolor[HTML]{000000}} \color[HTML]{F1F1F1} NA \\
\textbf{TEC-2} & {\cellcolor[HTML]{40B5C4}} \color[HTML]{F1F1F1} \ding{56} & {\cellcolor[HTML]{40B5C4}} \color[HTML]{F1F1F1} \ding{56} & {\cellcolor[HTML]{40B5C4}} \color[HTML]{F1F1F1} \ding{56} & {\cellcolor[HTML]{40B5C4}} \color[HTML]{F1F1F1} \ding{56} & {\cellcolor[HTML]{24419A}} \color[HTML]{F1F1F1} Graduated & {\cellcolor[HTML]{40B5C4}} \color[HTML]{F1F1F1} \ding{56} & {\cellcolor[HTML]{24419A}} \color[HTML]{F1F1F1} Graduated & {\cellcolor[HTML]{24419A}} \color[HTML]{F1F1F1} Graduated \\
\textbf{TEC-3} & {\cellcolor[HTML]{40B5C4}} \color[HTML]{F1F1F1} \ding{56} & {\cellcolor[HTML]{40B5C4}} \color[HTML]{F1F1F1} \ding{56} & {\cellcolor[HTML]{40B5C4}} \color[HTML]{F1F1F1} \ding{56} & {\cellcolor[HTML]{24419A}} \color[HTML]{F1F1F1} Graduated & {\cellcolor[HTML]{40B5C4}} \color[HTML]{F1F1F1} \ding{56} & {\cellcolor[HTML]{40B5C4}} \color[HTML]{F1F1F1} \ding{56} & {\cellcolor[HTML]{E1F3B2}} \color[HTML]{000000} Retired & {\cellcolor[HTML]{E1F3B2}} \color[HTML]{000000} Retired \\
\textbf{TEC-4} & {\cellcolor[HTML]{40B5C4}} \color[HTML]{F1F1F1} \ding{56} & {\cellcolor[HTML]{40B5C4}} \color[HTML]{F1F1F1} \ding{56} & {\cellcolor[HTML]{40B5C4}} \color[HTML]{F1F1F1} \ding{56} & {\cellcolor[HTML]{24419A}} \color[HTML]{F1F1F1} Graduated & {\cellcolor[HTML]{40B5C4}} \color[HTML]{F1F1F1} \ding{56} & {\cellcolor[HTML]{E1F3B2}} \color[HTML]{000000} Retired & {\cellcolor[HTML]{E1F3B2}} \color[HTML]{000000} Retired & {\cellcolor[HTML]{E1F3B2}} \color[HTML]{000000} Retired \\
\bottomrule
\end{tabularx}
} 
\egroup

%% file: tables_automatically_generated/table_h2.tex
\bgroup
\def\arraystretch{1.5}
\scalebox{0.714}{
\begin{tabularx}{1.39\textwidth}{
  p{25mm}
  >{\centering\arraybackslash}m{14mm}
  >{\centering\arraybackslash}m{20.5mm}
  >{\centering\arraybackslash}m{20.5mm}
  >{\centering\arraybackslash}m{20.5mm}
  >{\centering\arraybackslash}m{20.5mm}
  >{\centering\arraybackslash}m{20.5mm}
  >{\centering\arraybackslash}m{20.5mm}
  >{\centering\arraybackslash}m{20.5mm}
}
\toprule
 & \textbf{SWQ-1} & \textbf{SWQ-2.1} & \textbf{SWQ-2.6} & \textbf{SWQ-2.2} & \textbf{SWQ-2.3} & \textbf{SWQ-2.4} & \textbf{SWQ-2.5} & \textbf{SWQ-2.7} \\
\midrule
\textbf{COM-1} & {\cellcolor[HTML]{40B5C4}} \color[HTML]{F1F1F1} \ding{56} & {\cellcolor[HTML]{40B5C4}} \color[HTML]{F1F1F1} \ding{56} & {\cellcolor[HTML]{40B5C4}} \color[HTML]{F1F1F1} \ding{56} & {\cellcolor[HTML]{40B5C4}} \color[HTML]{F1F1F1} \ding{56} & {\cellcolor[HTML]{40B5C4}} \color[HTML]{F1F1F1} \ding{56} & {\cellcolor[HTML]{40B5C4}} \color[HTML]{F1F1F1} \ding{56} & {\cellcolor[HTML]{40B5C4}} \color[HTML]{F1F1F1} \ding{56} & {\cellcolor[HTML]{40B5C4}} \color[HTML]{F1F1F1} \ding{56} \\
\textbf{COM-2} & {\cellcolor[HTML]{24419A}} \color[HTML]{F1F1F1} Incubation & {\cellcolor[HTML]{40B5C4}} \color[HTML]{F1F1F1} \ding{56} & {\cellcolor[HTML]{40B5C4}} \color[HTML]{F1F1F1} \ding{56} & {\cellcolor[HTML]{40B5C4}} \color[HTML]{F1F1F1} \ding{56} & {\cellcolor[HTML]{E1F3B2}} \color[HTML]{000000} Post incubation & {\cellcolor[HTML]{E1F3B2}} \color[HTML]{000000} Post incubation & {\cellcolor[HTML]{E1F3B2}} \color[HTML]{000000} Post incubation & {\cellcolor[HTML]{E1F3B2}} \color[HTML]{000000} Post incubation \\
\textbf{POP-1} & {\cellcolor[HTML]{24419A}} \color[HTML]{F1F1F1} Incubation & {\cellcolor[HTML]{40B5C4}} \color[HTML]{F1F1F1} \ding{56} & {\cellcolor[HTML]{40B5C4}} \color[HTML]{F1F1F1} \ding{56} & {\cellcolor[HTML]{40B5C4}} \color[HTML]{F1F1F1} \ding{56} & {\cellcolor[HTML]{40B5C4}} \color[HTML]{F1F1F1} \ding{56} & {\cellcolor[HTML]{40B5C4}} \color[HTML]{F1F1F1} \ding{56} & {\cellcolor[HTML]{40B5C4}} \color[HTML]{F1F1F1} \ding{56} & {\cellcolor[HTML]{40B5C4}} \color[HTML]{F1F1F1} \ding{56} \\
\textbf{STA-1} & {\cellcolor[HTML]{40B5C4}} \color[HTML]{F1F1F1} \ding{56} & {\cellcolor[HTML]{E1F3B2}} \color[HTML]{000000} Post incubation & {\cellcolor[HTML]{40B5C4}} \color[HTML]{F1F1F1} \ding{56} & {\cellcolor[HTML]{E1F3B2}} \color[HTML]{000000} Post incubation & {\cellcolor[HTML]{40B5C4}} \color[HTML]{F1F1F1} \ding{56} & {\cellcolor[HTML]{E1F3B2}} \color[HTML]{000000} Post incubation & {\cellcolor[HTML]{E1F3B2}} \color[HTML]{000000} Post incubation & {\cellcolor[HTML]{E1F3B2}} \color[HTML]{000000} Post incubation \\
\textbf{STA-2} & {\cellcolor[HTML]{40B5C4}} \color[HTML]{F1F1F1} \ding{56} & {\cellcolor[HTML]{E1F3B2}} \color[HTML]{000000} Post incubation & {\cellcolor[HTML]{40B5C4}} \color[HTML]{F1F1F1} \ding{56} & {\cellcolor[HTML]{E1F3B2}} \color[HTML]{000000} Post incubation & {\cellcolor[HTML]{24419A}} \color[HTML]{F1F1F1} Incubation & {\cellcolor[HTML]{24419A}} \color[HTML]{F1F1F1} Incubation & {\cellcolor[HTML]{24419A}} \color[HTML]{F1F1F1} Incubation & {\cellcolor[HTML]{E1F3B2}} \color[HTML]{000000} Post incubation \\
\textbf{STA-3} & {\cellcolor[HTML]{24419A}} \color[HTML]{F1F1F1} Incubation & {\cellcolor[HTML]{40B5C4}} \color[HTML]{F1F1F1} \ding{56} & {\cellcolor[HTML]{40B5C4}} \color[HTML]{F1F1F1} \ding{56} & {\cellcolor[HTML]{40B5C4}} \color[HTML]{F1F1F1} \ding{56} & {\cellcolor[HTML]{40B5C4}} \color[HTML]{F1F1F1} \ding{56} & {\cellcolor[HTML]{40B5C4}} \color[HTML]{F1F1F1} \ding{56} & {\cellcolor[HTML]{40B5C4}} \color[HTML]{F1F1F1} \ding{56} & {\cellcolor[HTML]{40B5C4}} \color[HTML]{F1F1F1} \ding{56} \\
\textbf{STA-4} & {\cellcolor[HTML]{24419A}} \color[HTML]{F1F1F1} Incubation & {\cellcolor[HTML]{40B5C4}} \color[HTML]{F1F1F1} \ding{56} & {\cellcolor[HTML]{E1F3B2}} \color[HTML]{000000} Post incubation & {\cellcolor[HTML]{000000}} \color[HTML]{F1F1F1} NA & {\cellcolor[HTML]{000000}} \color[HTML]{F1F1F1} NA & {\cellcolor[HTML]{000000}} \color[HTML]{F1F1F1} NA & {\cellcolor[HTML]{000000}} \color[HTML]{F1F1F1} NA & {\cellcolor[HTML]{000000}} \color[HTML]{F1F1F1} NA \\
\textbf{STA-5} & {\cellcolor[HTML]{40B5C4}} \color[HTML]{F1F1F1} \ding{56} & {\cellcolor[HTML]{40B5C4}} \color[HTML]{F1F1F1} \ding{56} & {\cellcolor[HTML]{40B5C4}} \color[HTML]{F1F1F1} \ding{56} & {\cellcolor[HTML]{40B5C4}} \color[HTML]{F1F1F1} \ding{56} & {\cellcolor[HTML]{40B5C4}} \color[HTML]{F1F1F1} \ding{56} & {\cellcolor[HTML]{40B5C4}} \color[HTML]{F1F1F1} \ding{56} & {\cellcolor[HTML]{40B5C4}} \color[HTML]{F1F1F1} \ding{56} & {\cellcolor[HTML]{40B5C4}} \color[HTML]{F1F1F1} \ding{56} \\
\textbf{STA-6} {\tiny (dormant)} & {\cellcolor[HTML]{40B5C4}} \color[HTML]{F1F1F1} \ding{56} & {\cellcolor[HTML]{40B5C4}} \color[HTML]{F1F1F1} \ding{56} & {\cellcolor[HTML]{40B5C4}} \color[HTML]{F1F1F1} \ding{56} & {\cellcolor[HTML]{40B5C4}} \color[HTML]{F1F1F1} \ding{56} & {\cellcolor[HTML]{40B5C4}} \color[HTML]{F1F1F1} \ding{56} & {\cellcolor[HTML]{40B5C4}} \color[HTML]{F1F1F1} \ding{56} & {\cellcolor[HTML]{40B5C4}} \color[HTML]{F1F1F1} \ding{56} & {\cellcolor[HTML]{40B5C4}} \color[HTML]{F1F1F1} \ding{56} \\
\textbf{STA-6} {\tiny (non dormant)} & {\cellcolor[HTML]{40B5C4}} \color[HTML]{F1F1F1} \ding{56} & {\cellcolor[HTML]{40B5C4}} \color[HTML]{F1F1F1} \ding{56} & {\cellcolor[HTML]{40B5C4}} \color[HTML]{F1F1F1} \ding{56} & {\cellcolor[HTML]{40B5C4}} \color[HTML]{F1F1F1} \ding{56} & {\cellcolor[HTML]{40B5C4}} \color[HTML]{F1F1F1} \ding{56} & {\cellcolor[HTML]{40B5C4}} \color[HTML]{F1F1F1} \ding{56} & {\cellcolor[HTML]{40B5C4}} \color[HTML]{F1F1F1} \ding{56} & {\cellcolor[HTML]{40B5C4}} \color[HTML]{F1F1F1} \ding{56} \\
\textbf{STA-7} & {\cellcolor[HTML]{24419A}} \color[HTML]{F1F1F1} Incubation & {\cellcolor[HTML]{40B5C4}} \color[HTML]{F1F1F1} \ding{56} & {\cellcolor[HTML]{E1F3B2}} \color[HTML]{000000} Post incubation & {\cellcolor[HTML]{000000}} \color[HTML]{F1F1F1} NA & {\cellcolor[HTML]{000000}} \color[HTML]{F1F1F1} NA & {\cellcolor[HTML]{000000}} \color[HTML]{F1F1F1} NA & {\cellcolor[HTML]{000000}} \color[HTML]{F1F1F1} NA & {\cellcolor[HTML]{000000}} \color[HTML]{F1F1F1} NA \\
\textbf{STA-8} & {\cellcolor[HTML]{24419A}} \color[HTML]{F1F1F1} Incubation & {\cellcolor[HTML]{40B5C4}} \color[HTML]{F1F1F1} \ding{56} & {\cellcolor[HTML]{40B5C4}} \color[HTML]{F1F1F1} \ding{56} & {\cellcolor[HTML]{40B5C4}} \color[HTML]{F1F1F1} \ding{56} & {\cellcolor[HTML]{40B5C4}} \color[HTML]{F1F1F1} \ding{56} & {\cellcolor[HTML]{E1F3B2}} \color[HTML]{000000} Post incubation & {\cellcolor[HTML]{E1F3B2}} \color[HTML]{000000} Post incubation & {\cellcolor[HTML]{E1F3B2}} \color[HTML]{000000} Post incubation \\
\textbf{STA-9} & {\cellcolor[HTML]{24419A}} \color[HTML]{F1F1F1} Incubation & {\cellcolor[HTML]{24419A}} \color[HTML]{F1F1F1} Incubation & {\cellcolor[HTML]{40B5C4}} \color[HTML]{F1F1F1} \ding{56} & {\cellcolor[HTML]{E1F3B2}} \color[HTML]{000000} Post incubation & {\cellcolor[HTML]{40B5C4}} \color[HTML]{F1F1F1} \ding{56} & {\cellcolor[HTML]{E1F3B2}} \color[HTML]{000000} Post incubation & {\cellcolor[HTML]{E1F3B2}} \color[HTML]{000000} Post incubation & {\cellcolor[HTML]{24419A}} \color[HTML]{F1F1F1} Incubation \\
\textbf{TEC-1} & {\cellcolor[HTML]{40B5C4}} \color[HTML]{F1F1F1} \ding{56} & {\cellcolor[HTML]{40B5C4}} \color[HTML]{F1F1F1} \ding{56} & {\cellcolor[HTML]{40B5C4}} \color[HTML]{F1F1F1} \ding{56} & {\cellcolor[HTML]{000000}} \color[HTML]{F1F1F1} NA & {\cellcolor[HTML]{000000}} \color[HTML]{F1F1F1} NA & {\cellcolor[HTML]{000000}} \color[HTML]{F1F1F1} NA & {\cellcolor[HTML]{000000}} \color[HTML]{F1F1F1} NA & {\cellcolor[HTML]{000000}} \color[HTML]{F1F1F1} NA \\
\textbf{TEC-2} & {\cellcolor[HTML]{40B5C4}} \color[HTML]{F1F1F1} \ding{56} & {\cellcolor[HTML]{40B5C4}} \color[HTML]{F1F1F1} \ding{56} & {\cellcolor[HTML]{40B5C4}} \color[HTML]{F1F1F1} \ding{56} & {\cellcolor[HTML]{40B5C4}} \color[HTML]{F1F1F1} \ding{56} & {\cellcolor[HTML]{40B5C4}} \color[HTML]{F1F1F1} \ding{56} & {\cellcolor[HTML]{40B5C4}} \color[HTML]{F1F1F1} \ding{56} & {\cellcolor[HTML]{40B5C4}} \color[HTML]{F1F1F1} \ding{56} & {\cellcolor[HTML]{40B5C4}} \color[HTML]{F1F1F1} \ding{56} \\
\textbf{TEC-3} & {\cellcolor[HTML]{40B5C4}} \color[HTML]{F1F1F1} \ding{56} & {\cellcolor[HTML]{40B5C4}} \color[HTML]{F1F1F1} \ding{56} & {\cellcolor[HTML]{40B5C4}} \color[HTML]{F1F1F1} \ding{56} & {\cellcolor[HTML]{40B5C4}} \color[HTML]{F1F1F1} \ding{56} & {\cellcolor[HTML]{40B5C4}} \color[HTML]{F1F1F1} \ding{56} & {\cellcolor[HTML]{40B5C4}} \color[HTML]{F1F1F1} \ding{56} & {\cellcolor[HTML]{40B5C4}} \color[HTML]{F1F1F1} \ding{56} & {\cellcolor[HTML]{40B5C4}} \color[HTML]{F1F1F1} \ding{56} \\
\textbf{TEC-4} & {\cellcolor[HTML]{40B5C4}} \color[HTML]{F1F1F1} \ding{56} & {\cellcolor[HTML]{40B5C4}} \color[HTML]{F1F1F1} \ding{56} & {\cellcolor[HTML]{40B5C4}} \color[HTML]{F1F1F1} \ding{56} & {\cellcolor[HTML]{40B5C4}} \color[HTML]{F1F1F1} \ding{56} & {\cellcolor[HTML]{40B5C4}} \color[HTML]{F1F1F1} \ding{56} & {\cellcolor[HTML]{40B5C4}} \color[HTML]{F1F1F1} \ding{56} & {\cellcolor[HTML]{40B5C4}} \color[HTML]{F1F1F1} \ding{56} & {\cellcolor[HTML]{40B5C4}} \color[HTML]{F1F1F1} \ding{56} \\
\bottomrule
\end{tabularx}
} 
\egroup

%% file: tables_automatically_generated/table_h3.tex
\bgroup
\def\arraystretch{1.4}
\scalebox{0.9}{
\begin{tabularx}{1.1\textwidth}{
  p{25mm}
  >{\centering\arraybackslash}m{14mm}
  >{\centering\arraybackslash}m{14mm}
  >{\centering\arraybackslash}m{14mm}
  >{\centering\arraybackslash}m{14mm}
  >{\centering\arraybackslash}m{14mm}
  >{\centering\arraybackslash}m{14mm}
  >{\centering\arraybackslash}m{14mm}
  >{\centering\arraybackslash}m{14mm}
}
\toprule
 & \textbf{SWQ-1} & \textbf{SWQ-2.1} & \textbf{SWQ-2.6} & \textbf{SWQ-2.2} & \textbf{SWQ-2.3} & \textbf{SWQ-2.4} & \textbf{SWQ-2.5} & \textbf{SWQ-2.7} \\
\midrule
\textbf{COM-1} & {\cellcolor[HTML]{40B5C4}} \color[HTML]{F1F1F1} \ding{56} & {\cellcolor[HTML]{40B5C4}} \color[HTML]{F1F1F1} \ding{56} & {\cellcolor[HTML]{40B5C4}} \color[HTML]{F1F1F1} \ding{56} & {\cellcolor[HTML]{E1F3B2}} \color[HTML]{000000} Bypassed & {\cellcolor[HTML]{24419A}} \color[HTML]{F1F1F1} Graduated & {\cellcolor[HTML]{24419A}} \color[HTML]{F1F1F1} Graduated & {\cellcolor[HTML]{24419A}} \color[HTML]{F1F1F1} Graduated & {\cellcolor[HTML]{24419A}} \color[HTML]{F1F1F1} Graduated \\
\textbf{COM-2} & {\cellcolor[HTML]{40B5C4}} \color[HTML]{F1F1F1} \ding{56} & {\cellcolor[HTML]{40B5C4}} \color[HTML]{F1F1F1} \ding{56} & {\cellcolor[HTML]{40B5C4}} \color[HTML]{F1F1F1} \ding{56} & {\cellcolor[HTML]{24419A}} \color[HTML]{F1F1F1} Graduated & {\cellcolor[HTML]{40B5C4}} \color[HTML]{F1F1F1} \ding{56} & {\cellcolor[HTML]{E1F3B2}} \color[HTML]{000000} Bypassed & {\cellcolor[HTML]{E1F3B2}} \color[HTML]{000000} Bypassed & {\cellcolor[HTML]{E1F3B2}} \color[HTML]{000000} Bypassed \\
\textbf{POP-1} & {\cellcolor[HTML]{24419A}} \color[HTML]{F1F1F1} Graduated & {\cellcolor[HTML]{40B5C4}} \color[HTML]{F1F1F1} \ding{56} & {\cellcolor[HTML]{40B5C4}} \color[HTML]{F1F1F1} \ding{56} & {\cellcolor[HTML]{24419A}} \color[HTML]{F1F1F1} Graduated & {\cellcolor[HTML]{24419A}} \color[HTML]{F1F1F1} Graduated & {\cellcolor[HTML]{24419A}} \color[HTML]{F1F1F1} Graduated & {\cellcolor[HTML]{24419A}} \color[HTML]{F1F1F1} Graduated & {\cellcolor[HTML]{E1F3B2}} \color[HTML]{000000} Bypassed \\
\textbf{STA-1} & {\cellcolor[HTML]{40B5C4}} \color[HTML]{F1F1F1} \ding{56} & {\cellcolor[HTML]{40B5C4}} \color[HTML]{F1F1F1} \ding{56} & {\cellcolor[HTML]{40B5C4}} \color[HTML]{F1F1F1} \ding{56} & {\cellcolor[HTML]{E1F3B2}} \color[HTML]{000000} Bypassed & {\cellcolor[HTML]{E1F3B2}} \color[HTML]{000000} Bypassed & {\cellcolor[HTML]{E1F3B2}} \color[HTML]{000000} Bypassed & {\cellcolor[HTML]{E1F3B2}} \color[HTML]{000000} Bypassed & {\cellcolor[HTML]{E1F3B2}} \color[HTML]{000000} Bypassed \\
\textbf{STA-2} & {\cellcolor[HTML]{40B5C4}} \color[HTML]{F1F1F1} \ding{56} & {\cellcolor[HTML]{E1F3B2}} \color[HTML]{000000} Bypassed & {\cellcolor[HTML]{40B5C4}} \color[HTML]{F1F1F1} \ding{56} & {\cellcolor[HTML]{40B5C4}} \color[HTML]{F1F1F1} \ding{56} & {\cellcolor[HTML]{24419A}} \color[HTML]{F1F1F1} Graduated & {\cellcolor[HTML]{24419A}} \color[HTML]{F1F1F1} Graduated & {\cellcolor[HTML]{24419A}} \color[HTML]{F1F1F1} Graduated & {\cellcolor[HTML]{E1F3B2}} \color[HTML]{000000} Bypassed \\
\textbf{STA-3} & {\cellcolor[HTML]{24419A}} \color[HTML]{F1F1F1} Graduated & {\cellcolor[HTML]{40B5C4}} \color[HTML]{F1F1F1} \ding{56} & {\cellcolor[HTML]{40B5C4}} \color[HTML]{F1F1F1} \ding{56} & {\cellcolor[HTML]{24419A}} \color[HTML]{F1F1F1} Graduated & {\cellcolor[HTML]{24419A}} \color[HTML]{F1F1F1} Graduated & {\cellcolor[HTML]{24419A}} \color[HTML]{F1F1F1} Graduated & {\cellcolor[HTML]{24419A}} \color[HTML]{F1F1F1} Graduated & {\cellcolor[HTML]{E1F3B2}} \color[HTML]{000000} Bypassed \\
\textbf{STA-4} & {\cellcolor[HTML]{40B5C4}} \color[HTML]{F1F1F1} \ding{56} & {\cellcolor[HTML]{40B5C4}} \color[HTML]{F1F1F1} \ding{56} & {\cellcolor[HTML]{40B5C4}} \color[HTML]{F1F1F1} \ding{56} & {\cellcolor[HTML]{000000}} \color[HTML]{F1F1F1} NA & {\cellcolor[HTML]{000000}} \color[HTML]{F1F1F1} NA & {\cellcolor[HTML]{000000}} \color[HTML]{F1F1F1} NA & {\cellcolor[HTML]{000000}} \color[HTML]{F1F1F1} NA & {\cellcolor[HTML]{000000}} \color[HTML]{F1F1F1} NA \\
\textbf{STA-5} & {\cellcolor[HTML]{40B5C4}} \color[HTML]{F1F1F1} \ding{56} & {\cellcolor[HTML]{40B5C4}} \color[HTML]{F1F1F1} \ding{56} & {\cellcolor[HTML]{40B5C4}} \color[HTML]{F1F1F1} \ding{56} & {\cellcolor[HTML]{E1F3B2}} \color[HTML]{000000} Bypassed & {\cellcolor[HTML]{24419A}} \color[HTML]{F1F1F1} Graduated & {\cellcolor[HTML]{24419A}} \color[HTML]{F1F1F1} Graduated & {\cellcolor[HTML]{24419A}} \color[HTML]{F1F1F1} Graduated & {\cellcolor[HTML]{24419A}} \color[HTML]{F1F1F1} Graduated \\
\textbf{STA-6} {\tiny (dormant)} & {\cellcolor[HTML]{40B5C4}} \color[HTML]{F1F1F1} \ding{56} & {\cellcolor[HTML]{40B5C4}} \color[HTML]{F1F1F1} \ding{56} & {\cellcolor[HTML]{40B5C4}} \color[HTML]{F1F1F1} \ding{56} & {\cellcolor[HTML]{40B5C4}} \color[HTML]{F1F1F1} \ding{56} & {\cellcolor[HTML]{40B5C4}} \color[HTML]{F1F1F1} \ding{56} & {\cellcolor[HTML]{40B5C4}} \color[HTML]{F1F1F1} \ding{56} & {\cellcolor[HTML]{40B5C4}} \color[HTML]{F1F1F1} \ding{56} & {\cellcolor[HTML]{40B5C4}} \color[HTML]{F1F1F1} \ding{56} \\
\textbf{STA-6} {\tiny (non dormant)} & {\cellcolor[HTML]{40B5C4}} \color[HTML]{F1F1F1} \ding{56} & {\cellcolor[HTML]{40B5C4}} \color[HTML]{F1F1F1} \ding{56} & {\cellcolor[HTML]{40B5C4}} \color[HTML]{F1F1F1} \ding{56} & {\cellcolor[HTML]{40B5C4}} \color[HTML]{F1F1F1} \ding{56} & {\cellcolor[HTML]{40B5C4}} \color[HTML]{F1F1F1} \ding{56} & {\cellcolor[HTML]{40B5C4}} \color[HTML]{F1F1F1} \ding{56} & {\cellcolor[HTML]{40B5C4}} \color[HTML]{F1F1F1} \ding{56} & {\cellcolor[HTML]{40B5C4}} \color[HTML]{F1F1F1} \ding{56} \\
\textbf{STA-7} & {\cellcolor[HTML]{40B5C4}} \color[HTML]{F1F1F1} \ding{56} & {\cellcolor[HTML]{40B5C4}} \color[HTML]{F1F1F1} \ding{56} & {\cellcolor[HTML]{40B5C4}} \color[HTML]{F1F1F1} \ding{56} & {\cellcolor[HTML]{000000}} \color[HTML]{F1F1F1} NA & {\cellcolor[HTML]{000000}} \color[HTML]{F1F1F1} NA & {\cellcolor[HTML]{000000}} \color[HTML]{F1F1F1} NA & {\cellcolor[HTML]{000000}} \color[HTML]{F1F1F1} NA & {\cellcolor[HTML]{000000}} \color[HTML]{F1F1F1} NA \\
\textbf{STA-8} & {\cellcolor[HTML]{24419A}} \color[HTML]{F1F1F1} Graduated & {\cellcolor[HTML]{40B5C4}} \color[HTML]{F1F1F1} \ding{56} & {\cellcolor[HTML]{40B5C4}} \color[HTML]{F1F1F1} \ding{56} & {\cellcolor[HTML]{24419A}} \color[HTML]{F1F1F1} Graduated & {\cellcolor[HTML]{24419A}} \color[HTML]{F1F1F1} Graduated & {\cellcolor[HTML]{40B5C4}} \color[HTML]{F1F1F1} \ding{56} & {\cellcolor[HTML]{E1F3B2}} \color[HTML]{000000} Bypassed & {\cellcolor[HTML]{E1F3B2}} \color[HTML]{000000} Bypassed \\
\textbf{STA-9} & {\cellcolor[HTML]{24419A}} \color[HTML]{F1F1F1} Graduated & {\cellcolor[HTML]{40B5C4}} \color[HTML]{F1F1F1} \ding{56} & {\cellcolor[HTML]{40B5C4}} \color[HTML]{F1F1F1} \ding{56} & {\cellcolor[HTML]{40B5C4}} \color[HTML]{F1F1F1} \ding{56} & {\cellcolor[HTML]{24419A}} \color[HTML]{F1F1F1} Graduated & {\cellcolor[HTML]{40B5C4}} \color[HTML]{F1F1F1} \ding{56} & {\cellcolor[HTML]{E1F3B2}} \color[HTML]{000000} Bypassed & {\cellcolor[HTML]{E1F3B2}} \color[HTML]{000000} Bypassed \\
\textbf{TEC-1} & {\cellcolor[HTML]{40B5C4}} \color[HTML]{F1F1F1} \ding{56} & {\cellcolor[HTML]{40B5C4}} \color[HTML]{F1F1F1} \ding{56} & {\cellcolor[HTML]{40B5C4}} \color[HTML]{F1F1F1} \ding{56} & {\cellcolor[HTML]{000000}} \color[HTML]{F1F1F1} NA & {\cellcolor[HTML]{000000}} \color[HTML]{F1F1F1} NA & {\cellcolor[HTML]{000000}} \color[HTML]{F1F1F1} NA & {\cellcolor[HTML]{000000}} \color[HTML]{F1F1F1} NA & {\cellcolor[HTML]{000000}} \color[HTML]{F1F1F1} NA \\
\textbf{TEC-2} & {\cellcolor[HTML]{40B5C4}} \color[HTML]{F1F1F1} \ding{56} & {\cellcolor[HTML]{40B5C4}} \color[HTML]{F1F1F1} \ding{56} & {\cellcolor[HTML]{40B5C4}} \color[HTML]{F1F1F1} \ding{56} & {\cellcolor[HTML]{E1F3B2}} \color[HTML]{000000} Bypassed & {\cellcolor[HTML]{E1F3B2}} \color[HTML]{000000} Bypassed & {\cellcolor[HTML]{40B5C4}} \color[HTML]{F1F1F1} \ding{56} & {\cellcolor[HTML]{E1F3B2}} \color[HTML]{000000} Bypassed & {\cellcolor[HTML]{24419A}} \color[HTML]{F1F1F1} Graduated \\
\textbf{TEC-3} & {\cellcolor[HTML]{40B5C4}} \color[HTML]{F1F1F1} \ding{56} & {\cellcolor[HTML]{40B5C4}} \color[HTML]{F1F1F1} \ding{56} & {\cellcolor[HTML]{40B5C4}} \color[HTML]{F1F1F1} \ding{56} & {\cellcolor[HTML]{24419A}} \color[HTML]{F1F1F1} Graduated & {\cellcolor[HTML]{24419A}} \color[HTML]{F1F1F1} Graduated & {\cellcolor[HTML]{24419A}} \color[HTML]{F1F1F1} Graduated & {\cellcolor[HTML]{24419A}} \color[HTML]{F1F1F1} Graduated & {\cellcolor[HTML]{40B5C4}} \color[HTML]{F1F1F1} \ding{56} \\
\textbf{TEC-4} & {\cellcolor[HTML]{40B5C4}} \color[HTML]{F1F1F1} \ding{56} & {\cellcolor[HTML]{40B5C4}} \color[HTML]{F1F1F1} \ding{56} & {\cellcolor[HTML]{40B5C4}} \color[HTML]{F1F1F1} \ding{56} & {\cellcolor[HTML]{24419A}} \color[HTML]{F1F1F1} Graduated & {\cellcolor[HTML]{24419A}} \color[HTML]{F1F1F1} Graduated & {\cellcolor[HTML]{40B5C4}} \color[HTML]{F1F1F1} \ding{56} & {\cellcolor[HTML]{24419A}} \color[HTML]{F1F1F1} Graduated & {\cellcolor[HTML]{E1F3B2}} \color[HTML]{000000} Bypassed \\
\bottomrule
\end{tabularx}
} 
\egroup

%% file: tables_automatically_generated/table_h4_bypassed_retired.tex
\bgroup
\def\arraystretch{1.4}
\scalebox{0.9}{
\begin{tabularx}{1.1\textwidth}{
  p{25mm}
  >{\centering\arraybackslash}m{14mm}
  >{\centering\arraybackslash}m{14mm}
  >{\centering\arraybackslash}m{14mm}
  >{\centering\arraybackslash}m{14mm}
  >{\centering\arraybackslash}m{14mm}
  >{\centering\arraybackslash}m{14mm}
  >{\centering\arraybackslash}m{14mm}
  >{\centering\arraybackslash}m{14mm}
}
\toprule
 & \textbf{SWQ-1} & \textbf{SWQ-2.1} & \textbf{SWQ-2.6} & \textbf{SWQ-2.2} & \textbf{SWQ-2.3} & \textbf{SWQ-2.4} & \textbf{SWQ-2.5} & \textbf{SWQ-2.7} \\
\midrule
\textbf{COM-1} & {\cellcolor[HTML]{40B5C4}} \color[HTML]{F1F1F1} \ding{56} & {\cellcolor[HTML]{40B5C4}} \color[HTML]{F1F1F1} \ding{56} & {\cellcolor[HTML]{24419A}} \color[HTML]{F1F1F1} Bypassed & {\cellcolor[HTML]{24419A}} \color[HTML]{F1F1F1} Bypassed & {\cellcolor[HTML]{E1F3B2}} \color[HTML]{000000} Retired & {\cellcolor[HTML]{E1F3B2}} \color[HTML]{000000} Retired & {\cellcolor[HTML]{E1F3B2}} \color[HTML]{000000} Retired & {\cellcolor[HTML]{E1F3B2}} \color[HTML]{000000} Retired \\
\textbf{COM-2} & {\cellcolor[HTML]{40B5C4}} \color[HTML]{F1F1F1} \ding{56} & {\cellcolor[HTML]{40B5C4}} \color[HTML]{F1F1F1} \ding{56} & {\cellcolor[HTML]{40B5C4}} \color[HTML]{F1F1F1} \ding{56} & {\cellcolor[HTML]{24419A}} \color[HTML]{F1F1F1} Bypassed & {\cellcolor[HTML]{E1F3B2}} \color[HTML]{000000} Retired & {\cellcolor[HTML]{40B5C4}} \color[HTML]{F1F1F1} \ding{56} & {\cellcolor[HTML]{E1F3B2}} \color[HTML]{000000} Retired & {\cellcolor[HTML]{24419A}} \color[HTML]{F1F1F1} Bypassed \\
\textbf{POP-1} & {\cellcolor[HTML]{40B5C4}} \color[HTML]{F1F1F1} \ding{56} & {\cellcolor[HTML]{40B5C4}} \color[HTML]{F1F1F1} \ding{56} & {\cellcolor[HTML]{E1F3B2}} \color[HTML]{000000} Retired & {\cellcolor[HTML]{40B5C4}} \color[HTML]{F1F1F1} \ding{56} & {\cellcolor[HTML]{24419A}} \color[HTML]{F1F1F1} Bypassed & {\cellcolor[HTML]{40B5C4}} \color[HTML]{F1F1F1} \ding{56} & {\cellcolor[HTML]{E1F3B2}} \color[HTML]{000000} Retired & {\cellcolor[HTML]{24419A}} \color[HTML]{F1F1F1} Bypassed \\
\textbf{STA-1} & {\cellcolor[HTML]{40B5C4}} \color[HTML]{F1F1F1} \ding{56} & {\cellcolor[HTML]{40B5C4}} \color[HTML]{F1F1F1} \ding{56} & {\cellcolor[HTML]{40B5C4}} \color[HTML]{F1F1F1} \ding{56} & {\cellcolor[HTML]{24419A}} \color[HTML]{F1F1F1} Bypassed & {\cellcolor[HTML]{40B5C4}} \color[HTML]{F1F1F1} \ding{56} & {\cellcolor[HTML]{24419A}} \color[HTML]{F1F1F1} Bypassed & {\cellcolor[HTML]{24419A}} \color[HTML]{F1F1F1} Bypassed & {\cellcolor[HTML]{24419A}} \color[HTML]{F1F1F1} Bypassed \\
\textbf{STA-2} & {\cellcolor[HTML]{40B5C4}} \color[HTML]{F1F1F1} \ding{56} & {\cellcolor[HTML]{40B5C4}} \color[HTML]{F1F1F1} \ding{56} & {\cellcolor[HTML]{40B5C4}} \color[HTML]{F1F1F1} \ding{56} & {\cellcolor[HTML]{40B5C4}} \color[HTML]{F1F1F1} \ding{56} & {\cellcolor[HTML]{24419A}} \color[HTML]{F1F1F1} Bypassed & {\cellcolor[HTML]{24419A}} \color[HTML]{F1F1F1} Bypassed & {\cellcolor[HTML]{40B5C4}} \color[HTML]{F1F1F1} \ding{56} & {\cellcolor[HTML]{24419A}} \color[HTML]{F1F1F1} Bypassed \\
\textbf{STA-3} & {\cellcolor[HTML]{40B5C4}} \color[HTML]{F1F1F1} \ding{56} & {\cellcolor[HTML]{24419A}} \color[HTML]{F1F1F1} Bypassed & {\cellcolor[HTML]{40B5C4}} \color[HTML]{F1F1F1} \ding{56} & {\cellcolor[HTML]{24419A}} \color[HTML]{F1F1F1} Bypassed & {\cellcolor[HTML]{24419A}} \color[HTML]{F1F1F1} Bypassed & {\cellcolor[HTML]{24419A}} \color[HTML]{F1F1F1} Bypassed & {\cellcolor[HTML]{E1F3B2}} \color[HTML]{000000} Retired & {\cellcolor[HTML]{24419A}} \color[HTML]{F1F1F1} Bypassed \\
\textbf{STA-4} & {\cellcolor[HTML]{24419A}} \color[HTML]{F1F1F1} Bypassed & {\cellcolor[HTML]{40B5C4}} \color[HTML]{F1F1F1} \ding{56} & {\cellcolor[HTML]{40B5C4}} \color[HTML]{F1F1F1} \ding{56} & {\cellcolor[HTML]{000000}} \color[HTML]{F1F1F1} NA & {\cellcolor[HTML]{000000}} \color[HTML]{F1F1F1} NA & {\cellcolor[HTML]{000000}} \color[HTML]{F1F1F1} NA & {\cellcolor[HTML]{000000}} \color[HTML]{F1F1F1} NA & {\cellcolor[HTML]{000000}} \color[HTML]{F1F1F1} NA \\
\textbf{STA-5} & {\cellcolor[HTML]{40B5C4}} \color[HTML]{F1F1F1} \ding{56} & {\cellcolor[HTML]{40B5C4}} \color[HTML]{F1F1F1} \ding{56} & {\cellcolor[HTML]{40B5C4}} \color[HTML]{F1F1F1} \ding{56} & {\cellcolor[HTML]{24419A}} \color[HTML]{F1F1F1} Bypassed & {\cellcolor[HTML]{E1F3B2}} \color[HTML]{000000} Retired & {\cellcolor[HTML]{E1F3B2}} \color[HTML]{000000} Retired & {\cellcolor[HTML]{E1F3B2}} \color[HTML]{000000} Retired & {\cellcolor[HTML]{E1F3B2}} \color[HTML]{000000} Retired \\
\textbf{STA-6} {\tiny (dormant)} & {\cellcolor[HTML]{40B5C4}} \color[HTML]{F1F1F1} \ding{56} & {\cellcolor[HTML]{40B5C4}} \color[HTML]{F1F1F1} \ding{56} & {\cellcolor[HTML]{40B5C4}} \color[HTML]{F1F1F1} \ding{56} & {\cellcolor[HTML]{40B5C4}} \color[HTML]{F1F1F1} \ding{56} & {\cellcolor[HTML]{40B5C4}} \color[HTML]{F1F1F1} \ding{56} & {\cellcolor[HTML]{40B5C4}} \color[HTML]{F1F1F1} \ding{56} & {\cellcolor[HTML]{40B5C4}} \color[HTML]{F1F1F1} \ding{56} & {\cellcolor[HTML]{40B5C4}} \color[HTML]{F1F1F1} \ding{56} \\
\textbf{STA-6} {\tiny (non dormant)} & {\cellcolor[HTML]{40B5C4}} \color[HTML]{F1F1F1} \ding{56} & {\cellcolor[HTML]{40B5C4}} \color[HTML]{F1F1F1} \ding{56} & {\cellcolor[HTML]{40B5C4}} \color[HTML]{F1F1F1} \ding{56} & {\cellcolor[HTML]{40B5C4}} \color[HTML]{F1F1F1} \ding{56} & {\cellcolor[HTML]{40B5C4}} \color[HTML]{F1F1F1} \ding{56} & {\cellcolor[HTML]{40B5C4}} \color[HTML]{F1F1F1} \ding{56} & {\cellcolor[HTML]{40B5C4}} \color[HTML]{F1F1F1} \ding{56} & {\cellcolor[HTML]{40B5C4}} \color[HTML]{F1F1F1} \ding{56} \\
\textbf{STA-7} & {\cellcolor[HTML]{40B5C4}} \color[HTML]{F1F1F1} \ding{56} & {\cellcolor[HTML]{40B5C4}} \color[HTML]{F1F1F1} \ding{56} & {\cellcolor[HTML]{40B5C4}} \color[HTML]{F1F1F1} \ding{56} & {\cellcolor[HTML]{000000}} \color[HTML]{F1F1F1} NA & {\cellcolor[HTML]{000000}} \color[HTML]{F1F1F1} NA & {\cellcolor[HTML]{000000}} \color[HTML]{F1F1F1} NA & {\cellcolor[HTML]{000000}} \color[HTML]{F1F1F1} NA & {\cellcolor[HTML]{000000}} \color[HTML]{F1F1F1} NA \\
\textbf{STA-8} & {\cellcolor[HTML]{40B5C4}} \color[HTML]{F1F1F1} \ding{56} & {\cellcolor[HTML]{40B5C4}} \color[HTML]{F1F1F1} \ding{56} & {\cellcolor[HTML]{40B5C4}} \color[HTML]{F1F1F1} \ding{56} & {\cellcolor[HTML]{24419A}} \color[HTML]{F1F1F1} Bypassed & {\cellcolor[HTML]{24419A}} \color[HTML]{F1F1F1} Bypassed & {\cellcolor[HTML]{24419A}} \color[HTML]{F1F1F1} Bypassed & {\cellcolor[HTML]{E1F3B2}} \color[HTML]{000000} Retired & {\cellcolor[HTML]{24419A}} \color[HTML]{F1F1F1} Bypassed \\
\textbf{STA-9} & {\cellcolor[HTML]{40B5C4}} \color[HTML]{F1F1F1} \ding{56} & {\cellcolor[HTML]{40B5C4}} \color[HTML]{F1F1F1} \ding{56} & {\cellcolor[HTML]{40B5C4}} \color[HTML]{F1F1F1} \ding{56} & {\cellcolor[HTML]{24419A}} \color[HTML]{F1F1F1} Bypassed & {\cellcolor[HTML]{40B5C4}} \color[HTML]{F1F1F1} \ding{56} & {\cellcolor[HTML]{24419A}} \color[HTML]{F1F1F1} Bypassed & {\cellcolor[HTML]{E1F3B2}} \color[HTML]{000000} Retired & {\cellcolor[HTML]{24419A}} \color[HTML]{F1F1F1} Bypassed \\
\textbf{TEC-1} & {\cellcolor[HTML]{40B5C4}} \color[HTML]{F1F1F1} \ding{56} & {\cellcolor[HTML]{40B5C4}} \color[HTML]{F1F1F1} \ding{56} & {\cellcolor[HTML]{40B5C4}} \color[HTML]{F1F1F1} \ding{56} & {\cellcolor[HTML]{000000}} \color[HTML]{F1F1F1} NA & {\cellcolor[HTML]{000000}} \color[HTML]{F1F1F1} NA & {\cellcolor[HTML]{000000}} \color[HTML]{F1F1F1} NA & {\cellcolor[HTML]{000000}} \color[HTML]{F1F1F1} NA & {\cellcolor[HTML]{000000}} \color[HTML]{F1F1F1} NA \\
\textbf{TEC-2} & {\cellcolor[HTML]{40B5C4}} \color[HTML]{F1F1F1} \ding{56} & {\cellcolor[HTML]{40B5C4}} \color[HTML]{F1F1F1} \ding{56} & {\cellcolor[HTML]{40B5C4}} \color[HTML]{F1F1F1} \ding{56} & {\cellcolor[HTML]{24419A}} \color[HTML]{F1F1F1} Bypassed & {\cellcolor[HTML]{24419A}} \color[HTML]{F1F1F1} Bypassed & {\cellcolor[HTML]{24419A}} \color[HTML]{F1F1F1} Bypassed & {\cellcolor[HTML]{24419A}} \color[HTML]{F1F1F1} Bypassed & {\cellcolor[HTML]{24419A}} \color[HTML]{F1F1F1} Bypassed \\
\textbf{TEC-3} & {\cellcolor[HTML]{40B5C4}} \color[HTML]{F1F1F1} \ding{56} & {\cellcolor[HTML]{40B5C4}} \color[HTML]{F1F1F1} \ding{56} & {\cellcolor[HTML]{40B5C4}} \color[HTML]{F1F1F1} \ding{56} & {\cellcolor[HTML]{24419A}} \color[HTML]{F1F1F1} Bypassed & {\cellcolor[HTML]{40B5C4}} \color[HTML]{F1F1F1} \ding{56} & {\cellcolor[HTML]{40B5C4}} \color[HTML]{F1F1F1} \ding{56} & {\cellcolor[HTML]{E1F3B2}} \color[HTML]{000000} Retired & {\cellcolor[HTML]{E1F3B2}} \color[HTML]{000000} Retired \\
\textbf{TEC-4} & {\cellcolor[HTML]{40B5C4}} \color[HTML]{F1F1F1} \ding{56} & {\cellcolor[HTML]{40B5C4}} \color[HTML]{F1F1F1} \ding{56} & {\cellcolor[HTML]{40B5C4}} \color[HTML]{F1F1F1} \ding{56} & {\cellcolor[HTML]{24419A}} \color[HTML]{F1F1F1} Bypassed & {\cellcolor[HTML]{40B5C4}} \color[HTML]{F1F1F1} \ding{56} & {\cellcolor[HTML]{E1F3B2}} \color[HTML]{000000} Retired & {\cellcolor[HTML]{E1F3B2}} \color[HTML]{000000} Retired & {\cellcolor[HTML]{E1F3B2}} \color[HTML]{000000} Retired \\
\bottomrule
\end{tabularx}
} 
\egroup

%% file: discussion.tex
\section{Discussion and Implications}\label{sec:discussion}

We commence this section by synthesizing our findings within the extant literature (summarized in Sect. \ref{sec:related}) and our previous work, i.e., \cite{alami2024free}, after which we follow with a discussion of potential implications within and beyond ASF and its incubation strategy.

In our previous work, we found no evidence of a consistent positive or negative impact of our selected sustainability metrics on software quality \cite{alami2024free}. However, we treated all ASFI projects equally, regardless of their developmental trajectory, e.g., retired, graduated, during incubation, or post-incubation. This approach obscured the variations in the sustainability-SWQ relationship across different trajectories. 

This study accounts for the evolving and shifting nature of sustainability practices over time and developmental phases. By separately analyzing our projects population during incubation, post-incubation, and bypassed incubation phases, we captured both the temporal and developmental variations. In addition, the separation of graduated and retired accentuated both the effectiveness of ASF governance and the ability of the projects themselves to adhere to sustainability practices. Our findings show that project-specific factors may play a larger role than ASF's structure governance.

Our previous study aimed to establish a foundational understanding of the sustainability-SWQ relationship by systematically analyzing multiple sustainability indicators across ASFI projects. The focus was on assessing whether these indicators had a consistent impact across projects' lifetime irrespective of their developmental stages. The findings from our previous work \cite{alami2024free} provided insights that shaped our current trajectory-based approach. It provided a baseline. Once we observed that sustainability metrics did not show a clear overall effect, it became clear that a more granular trajectory-based approach was needed.

Our findings contextualize and extend prior research on the sustainability-software quality (SWQ) relationship in FOSS communities. Unlike St{\u{a}}nciulescu et al. \cite{stuanciulescu2022code}, who used graduation and retirement as proxies for sustainability, we conceptualized sustainability as a multi-dimensional construct. Our design acknowledges that ASF governance alone does not guarantee project longevity. For example, while \textbf{H1} holds, it also prompts us to reflect that during incubation, both project populations, graduated and retired, receive similar support and mentoring from ASF, yet exhibit different sustainability-SWQ relationships. This highlights that ASF governance alone does not guarantee sustainability, and project-specific factors also influence it.

Ghapanchi \cite{ghapanchi2015predicting} found direct links between defect fixing, feature addition, and sustainability; our findings concurred that SWQ in FOSS is shaped by growth (increased features and PRs). However, software quality is also influenced by complex social and cultural dynamics, as seen in the ROS community \cite{alami2018influencers}. Our findings on contributor turnover also support Foucault et al.'s findings \cite{foucault2015impact}, suggesting that an increase in turnover among graduated and bypassed effects on SWQ is positive but weaker in retired projects. Our results suggest that despite an increase in turnover, graduated and bypassed projects show resilience. Adopting knowledge transfer and continuity strategies is highly likely to contribute to this resilience. Additionally, while Wang et al. \cite{wang2020unveiling} found that knowledge concentration among elite contributors negatively impacts certain SWQ metrics, our results indicate that a similar observation is made when comparing graduated and retired, but that is not the case when comparing bypassed and retired. This is highly likely because the size of projects in both populations differs significantly. In large projects, high knowledge concentration creates bottlenecks and weakens quality assurance practices like code review due to scalability challenges and concentrated knowledge distribution. On the other hand, in small retired projects, a high concentration of knowledge ensures tight quality control, despite project sustainability issues.

Taken together, our findings highlight that structured incubation enhances the sustainability-SWQ relationship in graduated projects, reinforcing the role of governance in sustaining long-term quality. However, bypassed projects demonstrate alternative sustainability mechanisms that allow them to also maintain stronger sustainability-SWQ relationship. In contrast, retired projects, despite receiving ASF mentoring, exhibit weaker sustainability-SWQ relationships, suggesting that external governance and mentoring alone do not always ensure sustainability. Internal project dynamics, commitment, and adaptability play equally crucial roles. 

\subsection{Implications}

The results of \textbf{H1} may prompt us to reflect on why ASF's structured governance and mentoring enhance sustainability and subsequently software quality outcomes during incubation for graduated but not retired projects. This challenges the assumption that formal governance structures alone are sufficient to sustain open-source projects and suggests that internal project dynamics, and self-directed sustainability strategies matter equally. Governance and mentoring alone are not a sufficient conditions for sustainability; it must be complemented by other sustainability factors that retired projects seem lacking. However, there is scarcity in knowledge and guidelines on how to strategize sustainability in the context of FOSS. For example, when contributor turnover goes high, what strategies communities can implement to mitigate its impact? Similarly, how can projects ensure that knowledge is effectively transferred and distributed to prevent bottlenecks and avoid risks of attrition, to mitigate the risks associated with high knowledge concentration? \textbf{What strategies can empower FOSS communities to alleviate the impacts of declining sustainability indicators and how to monitor and mitigate their risks proactively?}

This opens avenues for further research on how FOSS communities can develop self-directed sustainability strategies to both mitigate the impact of declining sustainability indicators (e.g., turnover, engagement decline, or knowledge concentration) and proactively monitor these indicators to trigger timely interventions. These interventions could include automated contributor health monitoring, mentorship programs for onboarding and retention, rotating core maintainership roles to distribute knowledge, and early-warning mechanisms to detect and address sustainability risks before they affect the community health and its software quality. \textbf{Understanding how communities can strategize their sustainability remains an under explored question.}

Insights from the analysis of \textbf{H3} and \textbf{H4}, show that bypassed projects challenge the notion that incubation is the only viable sustainability pathway. Even though their sustainability-SWQ relationship does not outperform graduated projects, it does for the retired ones. This suggests that while structured governance fosters sustainability, self-regulated projects can develop their own mechanisms to sustain quality. This finding calls for further research to investigate how these communities managed to remain sustainable. Although sustainability can be highly contextual, we can still learn from comparative case studies. For example, \textbf{what self-directed sustainability strategies do bypassed and sustainable FOSS projects adopt to remain sustainable, and how do they compare to governance-driven approaches in incubated projects?} \textbf{What differentiates sustainable FOSS projects from struggling or abandoned ones in terms of sustainability strategies, governance structures, and software quality outcomes?} Investigating these questions will help us inform FOSS sustainability actionably.

Although our work does not layout actionable recommendations, this study answered a fundamental question, i.e., the FOSS sustainability and SWQ relationship. By investigating this question, we managed to reflect on the key variable of this study, i.e., FOSS sustainability and propose avenues for future research directions.

%% file: validity.tex
\section{Limitations \& Threats to validity}\label{sec:validity}

\paragraph{\textbf{Limitations: }} 

We adopted only 16 indicators from the Lin{\aa}ker et al. \cite{linaaker2022characterize} framework. We excluded themes and indicators not available in repository data such as ``finance'' and ``culture'' and/or not well-acknowledged in the literature. This is due to the constraints of MSR methods. However, we prioritized the most significant indicators, and we captured a representative and meaningful aspect of FOSS sustainability.

The scope and breadth of software quality encompasses a wider range of attributes than those we have included \cite{khomh2012exploratory}. Yet, our coverage extensively covers aspects of defect density and several code quality metrics. Software quality remains a complex and multi-dimensional construct, including performance, usability, reliability, and maintainability metrics \cite{alami2022scrum}. However, our coverage extend beyond a single metric and comprehensive compared to previous work, e.g., \cite{wang2012survival,foucault2015impact,stuanciulescu2022code}.

Our study is limited to ASF projects. This selection was driven by the ability of this ecosystem to capture trajectories and developmental phases we may not find in other projects populations. While our sample contains a large and diverse projects size (342 projects), our conclusions may not be transferable to other FOSS communities. However, this limitation opens an avenue for future research to extend this investigation beyond ASF population.

\paragraph{\textbf{Internal validity: }} 

We encountered data availability and quality issues during the mining activity. Not all ASF projects have well-maintained repositories. For example, not all projects have JIRA issues labeled; therefore, we could not to compute \textbf{SWQ-1} for some projects. This may have introduced bias in the selection process. Nevertheless, our sample size remains large and contains projects range, covering scope, maturity levels, and sustainability levels.

Another potential threat to validity in our study is the confounding effect of project age in the analysis of sustainability metrics, particularly for bypassed projects. Bypassed projects tend to be older than both graduated and retired projects, which may introduce a bias in interpreting their sustainability indicators. In addition, they may also have  accumulate technical debt compared to younger project (i.e., graduated and retired). Subsequently, this debt was reflected in the computation of their code quality metrics.

We selected standard non-informative priors; i.e., priors that weigh all parameter values equally (no matter how unlikely they are). 
These priors may over-represent the probability of parameter values that are implausible.
However, this type of priors are common choice when no prior information is available (e.g.,~\cite{mcelreath2020}), and their only impact is an increase in uncertainty.
Our posterior plots do not exhibit large high density intervals (HDIs), which indicates that our priors did not introduce large uncertainty on the outcome of the analysis.

Our decision criteria depend on the units for each metric.
This choice was made to have consistent criteria for all our analyses. 
However, cases where we report stronger/weaker impact may effectively exhibit minor significance in practice.

Our analysis considers single predictor models. 
They quantify the impact of a single sustainability metric impacts a single quality metric, and compares it for different project types. 
However, it is possible that combinations of sustainability metrics have an impact while each of them individually do not. 
To study these effects, it is necessary to build multiple predictor models. 
Unfortunately, these models introduce complications in the interpretability of the results.
With multiple predictors, each predictor is to be interpreted conditioned on the other predictors in the model (e.g.,~\cite{mcelreath2020}).
On the contrary, our models have a simple and direct interpretation (see Sect.~\ref{sec:methods}).
This work aims at building a solid and easy-to-interpret foundation in the understanding of the interplay between sustainability and quality metrics for different project types. 
For these reasons, we have considered single predictor models in this work; and we leave as future work the study of combination of sustainability metrics on software quality.

\paragraph{\textbf{External validity: }} Outside the context of ASF, communities may vary in their approaches to sustainability, especially niche projects. Hence, we do not claim the generalization of our findings beyond our sample. However, the inclusion of bypassed projects mitigate to some extent this threat. This segment of our sample did not went through incubation and have developed different pathways to sustainability.

Part of our data quality filtering, we excluded projects without PRs in GitHub or Jira, empty issue trackers, etc. This exercise may have caused a selection bias threat. The final sample could be over-representative of projects with consistent configuration of Github and JIRA, for example, favoring projects with well-structured software development infrastructure.

%% file: conclusion.tex
\section{Conclusion}\label{sec:conclusion}
 
In this study, we sought to investigate the relationship between FOSS sustainability and software quality and how the developmental phases, retired, graduated during incubation, graduated post-incubation, and bypassed incubation, moderate this relationship. We used 16 sustainability metrics seven software quality metrics to investigate this relationship. 

In conclusion, we find that selected sustainability metrics exhibit distinct relationships with software quality across different project trajectories. Graduated projects demonstrated the strongest sustainability-SWQ relationship, both during and post-incubation, suggesting that structured governance and mentorship are effective strategies to promote sustainability. In contrast, retired projects exhibited weaker sustainability-SWQ relationships, despite receiving similar governance support, highlighting that formal oversight alone is not sufficient for sustaining projects. Bypassed projects, while not outperforming graduated ones, showed comparable sustainability-SWQ relationships.

We conclude that while structured governance and mentorship in the incubation process reinforce sustainability practices and their impact on software quality is stronger, they are not a guarantee as universal sustainability strategies, as seen in retired projects. Similarly, bypassed projects demonstrate that self-regulated sustainability strategies can be also effective, but they do not necessarily surpass the benefits of structured incubation.

Our study advances our understanding of the sustainability-SWQ relationship, and it also raises questions about how to advance the empirical software engineering community contribution in advancing FOSS sustainability. For example, we should investigate how communities can proactively manage sustainability challenges, develop self-directed resilience mechanisms, and optimize governance strategies to sustain both community health and subsequently enhanced software quality.

%% file: acks.tex
\section*{Acknowledgments}\label{sec:acks}

\noindent This study was partly funded by STIBOFONDEN.